\documentclass[%
 reprint,
 amsmath,amssymb,
 aps,
 pre,
]{revtex4-2}

\usepackage{graphicx}
\usepackage{dcolumn}
\usepackage{bm}

\usepackage{color}

\DeclareTextSymbolDefault{\textquotedbl}{T1}
\providecommand{\tabularnewline}{\\}

\usepackage{xcolor}

\begin{document}
\title{Unconventional low temperature features in the one-dimensional frustrated $q$-state Potts model}

\author{Yury Panov}
\affiliation{Ural Federal University, 19 Mira street, 620002 Ekaterinburg, Russia}

\author{Onofre Rojas}
\affiliation{Department of Physics, Federal University of Lavras, 37200-900 Lavras-MG, Brazil}

\date{\today}

\begin{abstract}
Here we consider a one-dimensional $q$-state Potts model with an
external magnetic field and an anisotropic interaction that selects
neighboring sites that are in the spin state 1. The present model
exhibits an unusual behavior in the low-temperature region, where
we observe an anomalous vigorous change in the entropy for a given
temperature. There is a steep behavior at a given temperature in entropy
as a function of temperature, quite similar to first-order discontinuity,
but there is no jump in the entropy. Similarly, second derivative
quantities like specific heat and magnetic susceptibility also exhibit
a strong acute peak rather similar to second-order phase transition
divergence, but once again there is no singularity at this point.
Correlation length also confirms this anomalous behavior at the same
given temperature, showing a strong and sharp peak which easily one
may confuse with a divergence. The temperature where occurs this anomalous
feature we call pseudo-critical temperature. We have analyzed physical
quantities, like correlation length, entropy, magnetization, specific
heat, magnetic susceptibility, and distant pair correlation functions.
Furthermore, we analyze the pseudo-critical exponent that satisfy
a class of universality previously identified in the literature for
other one-dimensional models, these pseudo-critical exponents are:
for correlation length $\nu=1$, specific heat $\alpha=3$ and magnetic
susceptibility $\mu=3$.
\end{abstract}


\maketitle

\section{Introduction}

The advantage of exactly solvable models is its easy handling to analyze
several properties, which can show interesting features despite it
is simplicity. In contrast, more detailed models are rarely exactly
solvable, and it would restrict us to performing only through numerical
computations, which prevents further analysis of these types of models.
Some one-dimensional models\cite{baxter} can help us understand
and predict leading behavior in more complex models. From the experimental
side, one-dimensional models accurately describe several chemical
compounds\cite{Ghulghazaryan,Hovhannisyan}. That is why the one-dimensional
models are quite important to investigate, both from theoretical and
experimental points of view.

Earlier in the fifties, van Hove\cite{Hove} proposed a theorem to
verify the absence of phase transition in uniform one-dimensional
models with short-range interaction. The validity of the proposed
theorem follows the conditions: (i) homogeneity, (ii) the Hamiltonian
should not include particles positions terms (like external fields).
(iii) hard-core particles. Based on the Perron-Frobenius theorem \cite{Ninio},
condition, the van Hove theorem\cite{Hove} is restricted to limited
one-dimensional systems. Later, Cuesta and Sanchez\cite{cuesta}
tried to extend the theorem of non-existence phase transition for
a more general one-dimensional system. Mainly, they included an external
field and considered point-like particles, which extends the theorem.
Even with this extension, it is still far from being a fully general
theorem of non-existence phase transition. 

There are unusual one-dimensional models with a short-range coupling
that exhibit a phase transition at finite temperature. The zipper
or Kittel model \cite{kittel}, which is one of the simplest models
with a finite size transfer matrix that exhibits a first-order phase
transition. 
Another model, considered by Chui-Wicks\cite{chui}, is typical of models called solid-on-solid for surface growth. 
It has the infinite dimension transfer-matrix and is exactly solvable. 
Because of the impenetrable condition of the subtract, the model shows
the existence of a finite temperature phase transition. 
One more model is that considered by Dauxois-Peyrard\cite{dauxois}, with an infinite
dimension transfer matrix, which can be explored numerically. Lately,
Sarkanych et al.\cite{sarkanych} proposed a one-dimensional Potts
model with invisible states and short-range coupling. The term invisible
means an additional energy degeneracy, which only contributes to the
entropy, but not the interaction energy. They named these states the
invisible states, which generate the first-order phase transition.

Motivated by low-dimensional systems, such as the simple zipper model\cite{kittel}
that describes the long-chain nucleotides of deoxyribonucleic acid
(DNA), Zimm and Bragg\cite{zimm-bragg} introduced an essentially
phenomenological cooperative parameter, which provides narrow helix-coil
transitions. Since that several investigations were driven in the
literature\cite{Badasyan10,ananikyan,Badasyan13,tonoyan}. Cooperative
systems in one-dimension can be well represented by Potts-like models\cite{Badasyan10,Badasyan13},
where the helix-coil transition in polypeptides\cite{ananikyan}
can be studied, which is a typical application of theoretical physics
to macromolecular systems, the results of which are quite appropriate
to understand the physical properties of the helix-coil transition.
The polycyclic aromatic surface elements of the carbon nanotube (CNT)
and the aromatic DNA provide reversible adsorption. Tonoyan et al.\cite{tonoyan}
adapted the Hamiltonian of the zipper model in order to take into
account the DNA-CNT interactions.

Potts model is a generalization of the Ising model to more than two
components, such as interacting spins in a crystalline lattice. Standard
$q$-state Potts model \cite{Wu}, with $q\geqslant2$ has been assumed
as an integer denoting the number of states of each site. Potts model
is quite relevant in statistical physics. In some crystal-lattices
would occur vacancies, which leads to a site diluted Potts model.
The site dilute $q$-state Potts model\cite{aharony,lubensky} is equivalent
to $q+1$-state standard Potts model\cite{Wu}. Chaves and Riera\cite{chaves}
investigated a particular case of dilute Potts chain. Recently, a
different dilute Ising spin-1 chain\cite{panov} was also studied
in the framework of the projection operator.

An unusual property called pseudo-transition was observed in some
recent works: Like in a double-tetrahedral chain of localized Ising
spins with mobile electrons showing a strong thermal excitation that
easily suggests the existence of a first-order phase transition\cite{Galisova,galisova17,*galisova20}.
Similarly, the frustrated spin-1/2, Ising-Heisenberg's three-leg tube
exhibited a pseudo-transition\cite{strk-cav}. In reference \cite{on-strk},
this property was also observed when studying the specific heat, where
reported a sharp peak on the spin-1/2 Ising-Heisenberg ladder with
alternating Ising and Heisenberg inter-leg couplings. This weird property
was even observed in the spin-1/2 Ising diamond chain in the neighboring
of the pseudo-transition\cite{psd-Ising}. Besides, deeper investigations
were performed on this peculiar property in \cite{pseudo}. Additionally,
the distant correlation functions have been studied around pseudo-transition
for a spin-1/2 Ising-XYZ diamond chain\cite{Isaac}. A bit different
proposal to identify the pseudo-transition in one-dimensional models\cite{ph-bd,*Rojas2020}
was explored in the framework of the phase boundary residual entropy
relationship with the finite temperature pseudo-transition. Further
investigation around the pseudo-transition was also focused on the
universality and pseudo-critical exponents of one-dimensional models\cite{unv-cr-exp}.

We organize the article as follows. In Section~\ref{sec:Potts_chain} we present the model
and analyze the zero-temperature phase diagrams. In Section~\ref{sec:Thermodynamics}, 
we study the thermodynamics of the model and explore an anomalous phenomenon called pseudo-transition 
for several physical quantities. In Section~\ref{sec:PDF}, 
we investigate the manifest of the pseudo-transition 
in terms of the distant pair correlation functions and correlation length. 
The pseudo critical exponents for the correlation length, specific heat, 
and susceptibility we discuss in Section~\ref{sec:critical_exp}. 
Finally, in Section~\ref{sec:Conclusion}, we present our conclusions.

\section{\label{sec:Potts_chain}The frustrated Potts chain}

Let us consider a Potts model\cite{Wu}, here we
assume the one-dimensional case, whose Hamiltonian becomes 
\begin{equation}
H=-\sum_{i=1}^{N}\left\{ J\delta_{\sigma_{i},\sigma_{i+1}}+K\delta_{\sigma_{i},1}\delta_{1,\sigma_{i+1}}+h\delta_{\sigma_{i},1}\right\} ,
\label{eq:PottsChain1}
\end{equation}
with $\sigma=\left\{ 1,\ldots q\right\} $. Whereas $J$ is bound
coupling parameter, $h$ denotes the external field aligned to state
1, and $K$ denotes the parameter of an anisotropic interaction that selects
neighboring sites that are in the spin state 1. 

The Hamiltonian \eqref{eq:PottsChain1} for the case $q=2$, like
standard Potts model drops into spin-1/2 Ising chain model.
For the case of $q\geqslant3$, the model becomes a frustrated system for certain choices of parameters, as shown below.

It is worth notice that the Hamiltonian \eqref{eq:PottsChain1}
can also be equivalent to a diluted Potts model with $q-1$ states,
as demonstrated by Wu~\cite{Wu}. 
The features of the critical properties of the two-dimensional diluted Potts model were studied earlier using Fortuin-Kasteleyn clusters~\cite{Janke2004} and the transfer matrix method~\cite{Qian2005}.

\subsection{Zero temperature phase diagram}

In order to analyze the phase diagram of $q$-state Potts model at
zero temperature, we identify four ground states assuming $q\geqslant3$
for the model \eqref{eq:PottsChain1}, which read below 

\begin{alignat}{1}
|FM_{1}\rangle= & \prod_{i=1}^{N}\bigl(|1\rangle_{i}\bigr), \label{eq:wfFM1}\\
|FM_{2}\rangle= & \prod_{i=1}^{N}\bigl(|j\rangle_{i}\bigr), \quad j=\{2,\dots,q\}, \label{eq:wfFM2}\\
|FR_{1}\rangle= & \prod_{i=1}^{N/2}\left(|1\rangle_{2i-1}|\sigma_{i}\rangle_{2i}\right), \label{eq:wfFR1}\\
|FR_{2}\rangle= & \prod_{i=1}^{N}\left(|\sigma_{i}\rangle\right),\quad\sigma_{i}\neq\sigma_{i\pm1},  \label{eq:wfFR2}
\end{alignat}
where $\sigma_{i}=\{2,\dots,q\}$.
Additional information, concerning phases and phase-boundaries, are
listed in Table~\ref{tab:mgs1} for different physical quantities
at zero temperature, like magnetization, entropy, and pair distribution
functions (PDF), which can be obtained by taking the limit of $T\rightarrow0$
of the quantities \eqref{eq:s}, \eqref{eq:m} and  \eqref{eq:pdf}.

\begin{figure*}
\includegraphics[width=0.3\textwidth]{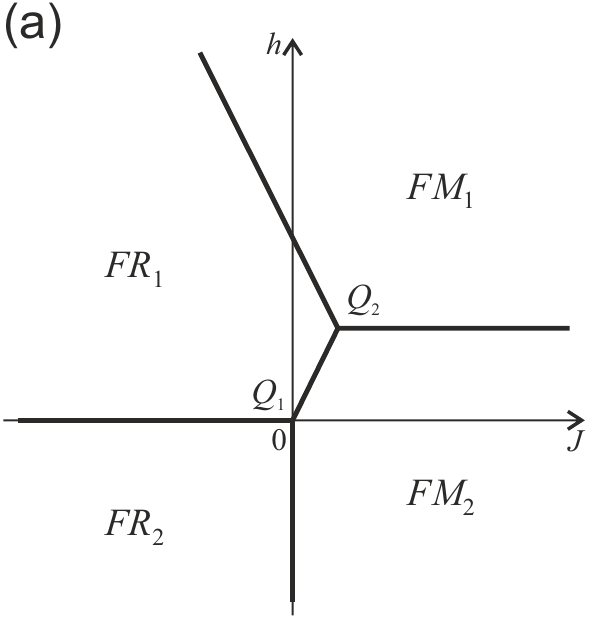} 
\quad{}\includegraphics[width=0.3\textwidth]{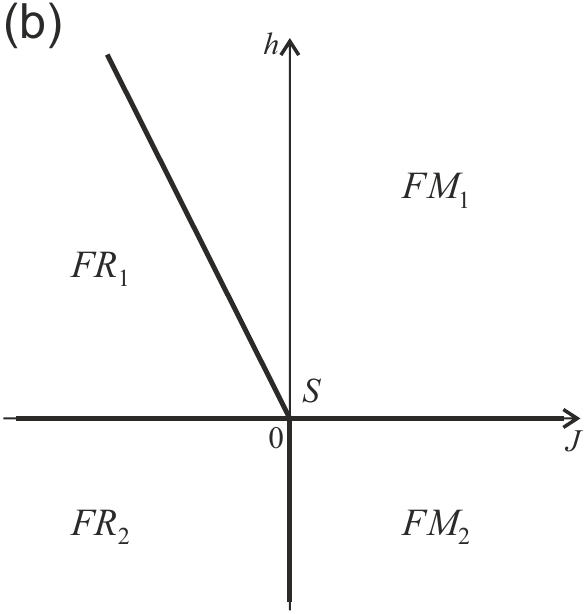}
\quad{}\includegraphics[width=0.3\textwidth]{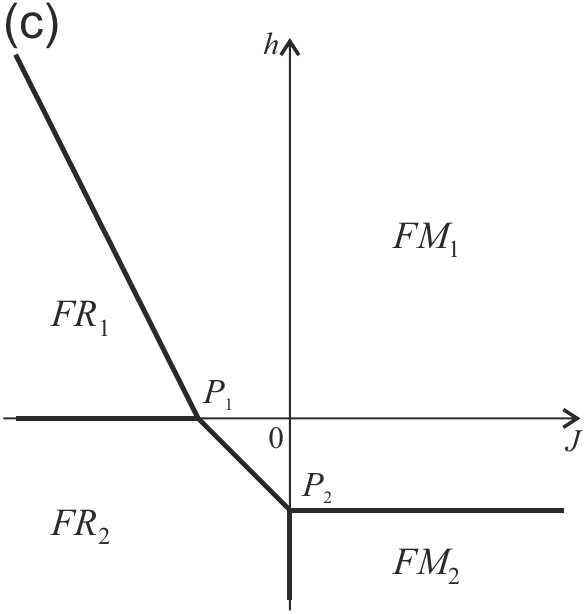} 
\caption{The ground state phase diagrams of the $q$-state 
Potts chain with Hamiltonian (\ref{eq:PottsChain1}) for
$q\geqslant3$. (a) $K<0$, (b) $K=0$, and (c) $K>0$. Here the
points are: $Q_{1}(0,0)$, $Q_{2}(-K/2,-K)$, $S(0,0)$, $P_{1}(-K,0)$,
$P_{2}(0,-K)$.}
\label{fig:PDchain1}
\end{figure*}

In Fig.~\ref{fig:PDchain1}a is illustrated schematically the ground
state phase diagram in the plane $J-h$, for the Hamiltonian \eqref{eq:PottsChain1}
assuming $K<0$. Here we observe four phases illustrated above. Similarly,
in panel (b) and (c) are illustrated alternative phase diagrams for
$K=0$ and $K>0$, respectively.

The first state is a type of ferromagnetic ($FM_{1}$) phase where
all sites are in state $\sigma_{i}=1$, with energy per spin $E_{FM_{1}}=-(J+K+h)$.
The second state is another type of \textquotedbl ferromagnetic\textquotedbl{} ($FM_{2})$ phase  
with energy $E_{FM_{2}}=-J$. 
The average fraction of pairs of adjacent spins in the same state $\mu$, where $\mu=2,\ldots q$, 
is PDF $g_{\mu,\mu}^{(1)} = 1/(q-1)$ (see Table 1). 
At the same time, $g_{\mu,\mu'}^{(1)} = 0$, where $\mu\neq\mu'$, 
which means that in the thermodynamic limit, 
the fraction of pairs with different spin states and their contribution to the energy of the system are equal to zero. 
This implies that, in the general case, the $FM_{2}$ phase consists of $q-1$ sorts 
of equivalent macroscopic ferromagnetic domains with spins in the $\mu$ state. 
Equation~\eqref{eq:wfFM2} corresponds to the single-domain case. 
The entropy of both $FM_{1}$ and $FM_{2}$ phases is zero. The third
is a type frustrated ($FR_{1}$) phase, with alternating sites are
in states $\sigma_{2i-1}=1$, while the other sites $\sigma_{2i}$
can take $\sigma_{2i}=2,\ldots q$, and the corresponding energy is
$E_{FR_{1}}=-h/2$. This phase state is frustrated at $q>2$. Due
to every second site can be in any of the $q-1$ states, the entropy per spin 
is equal to $\mathcal{S}=\frac{1}{2}\ln(q-1)$. The fourth state is
another type of frustrated ($FR_{2}$) phase, in each site $\sigma_{i}$
can take independently $\sigma_{i}=\{2,\dots,q\}$ but $\sigma_{i}\ne\sigma_{i\pm1}$,
whose corresponding energy is $E_{FR_{2}}=0$. In this case, each
subsequent site of the chain can be in one of $q-2$ states, and the
entropy is equal to $\mathcal{S}=\ln(q-2)$. If $q>3$, the $FR_{2}$
phase is evidently frustrated state. Although, for $q=3$, there is
an alternation of sites in states $\sigma_{i}=2$ and $\sigma_{i}=3$,
and the entropy of the $FR_{2}$ phase is zero.

It is important to note that two different situations can occur at
phase boundaries. The first case is when the states of two adjacent
phases are mixed at the microscopic level. For example, for the $FR_{1}$-$FM_{1}$
boundary, the $FR_{1}$ state of any pair of sites can be changed
to the $FM_{1}$ state, and vice versa. Such a replacement does not
lead to the appearance of microscopic states from other phases, and
the energy of the system does not change. As a result, the entropy
of such a mixed state at the phase boundary is greater than the entropy
of the adjacent phases. A similar situation is observed for the boundaries
$FR_{1}$-$FR_{2}$, $FR_{2}$-$FM_{2}$, and $FR_{1}$-$FM_{2}$
at $K<0$. In the second case, it can be a pure state of one of the
adjacent phases, or the phase separation, when each phase is represented
by the macroscopic domains. So, on the $FR_{2}$-$FM_{1}$ boundary
for $K>0$, the replacement for a pair of neighboring nodes in state
$FR_{2}$ by state $FM_{1}$ leads to the appearance of $FR_{1}$ 
states, which are energetically unfavorable. A similar situation occurs
at the $FM_{1}$-$FM_{2}$ boundary. Note that in the $FM_{1}$-$FM_{2}$ interface curve
the residual entropy is zero.

\section{\label{sec:Thermodynamics}Thermodynamics}

The frustrated $q$-state Potts model Hamiltonian \eqref{eq:PottsChain1} 
can be solved through transfer matrix technique, which results in
a $q$-dimensional matrix, given by 
\begin{equation}
\mathbf{V}=\begin{pmatrix}d_{1} & t_{1} & t_{1} & \cdots & t_{1} & t_{1}\\
t_{1} & d_{2} & t_{2} & \cdots & t_{2} & t_{2}\\
t_{1} & t_{2} & d_{2} & \cdots & t_{2} & t_{2}\\
\vdots & \vdots & \vdots & \ddots & \vdots & \vdots\\
t_{1} & t_{2} & t_{2} & \cdots & d_{2} & t_{2}\\
t_{1} & t_{2} & t_{2} & \cdots & t_{2} & d_{2}
\end{pmatrix},
\end{equation}
where $d_{1}=xkz$, $d_{2}=x$, $t_{1}=\sqrt{z}$, $t_{2}=1$, and
$x=\mathrm{e}^{\beta J}$, $k=\mathrm{e}^{\beta K}$, $z=\mathrm{e}^{\beta h}$. 

Let us write the transfer matrix eigenvalues similarly to that defined
in reference \cite{pseudo}, that being so, we have

\begin{alignat}{1}
\lambda_{1}= &\, \frac{1}{2}\left(w_{1}+w_{-1}+\sqrt{(w_{1}-w_{-1})^{2}+4w_{0}^{2}}\right),\label{eq:Lmbd1}\\
\lambda_{2}= &\, \frac{1}{2}\left(w_{1}+w_{-1}-\sqrt{(w_{1}-w_{-1})^{2}+4w_{0}^{2}}\right),\label{eq:Lmbd2}\\
\lambda_{j}= &\, d_{2}-t_{2},\quad j=\{3,4,\dots,q\},\label{eq:Lmbdj}
\end{alignat}
where 
\begin{eqnarray}
w_{1} &=& d_{1},\\
w_{-1} &=& d_{2}+\left(q-2\right)t_{2},\\
w_{0} &=& \sqrt{q-1}\,t_{1}.
\end{eqnarray}

The corresponding transfer matrix eigenvectors are 
\begin{alignat}{1}
|u_{1}\rangle= & \cos(\phi)|1\rangle+\tfrac{\sin(\phi)}{\sqrt{q-1}}\sum_{\mu=2}^{q}|\mu\rangle,\label{eq:u1}\\
|u_{2}\rangle= & -\sin(\phi)|1\rangle+\tfrac{\cos(\phi)}{\sqrt{q-1}}\sum_{\mu=2}^{q}|\mu\rangle,\label{eq:u2}\\
|u_{j}\rangle= & \sqrt{\tfrac{j-2}{j-1}}\Bigl(\tfrac{1}{j-2}\sum_{\mu=2}^{j-1}|\mu\rangle-|j\rangle\Bigr),\quad j=\{3,\cdots,q\},\label{eq:u3}
\end{alignat}
where $\phi=\frac{1}{2}\cot^{-1}\left(\frac{w_{1}-w_{-1}}{2w_{0}}\right)$,
with $-\frac{\pi}{4}\leqslant\phi\leqslant\frac{\pi}{4}$.

By using the transfer matrix eigenvalues, we express the partition
function as follows

\begin{alignat}{1}
Z_{N}= &\, \lambda_{1}^{N}+\lambda_{2}^{N}+(q-2)\lambda_{3}^{N}\nonumber \\
= &\, \lambda_{1}^{N}\left\{ 1+\Bigl(\tfrac{\lambda_{2}}{\lambda_{1}}\Bigr)^{N}+(q-2)\Bigl(\tfrac{\lambda_{3}}{\lambda_{1}}\Bigr)^{N}\right\} .\label{eq:Zn}
\end{alignat}
It is evident that the eigenvalues satisfy the following relation
$\lambda_{1}>\lambda_{2}\geqslant\lambda_{3}$. Hence, assuming $q$
finite, the free energy per spin in thermodynamic limit reduces to 

\begin{equation}
f=-T\ln\left(\lambda_{1}\right).
\label{eq:FE}
\end{equation}

Note that the free energy is a continuous function with no singularity
or discontinuity, thus we do not expect any real phase transition at
finite temperature.

\subsection{Pseudo-Critical temperature}

Recently pseudo-critical temperature has been discussed in Ising and
Ising-Heisenberg spin models\cite{Galisova,strk-cav,on-strk,pseudo}, 
in several one-dimensional spin models.

To find pseudo-critical temperature, we follow the same strategy to
that used in reference\cite{pseudo}. In our case the largest eigenvalues
has the same structure to that found in reference\cite{pseudo}, 
so necessary conditions for a pseudo-transition are met, if 
$w_1 \sim w_{-1} \ggg w_0$, $|w_1 - w_{-1}| \gg w_0$. 
The pseudo-transition point can be obtained when the first term inside
the square root of $\lambda_{1}$ given by eq.\eqref{eq:Lmbd1} turns to zero,
which gives
\begin{equation}
	e^{\frac{J+K+h}{T_p}} = q-2 + e^{\frac{J}{T_p}} .
\label{eq:psd-eq}
\end{equation}
In principle, using the above relation, one can find the critical
temperature as a function of some Hamiltonian parameters. 

The $q$-state Potts chain does not exhibit a real spontaneous long
range order at finite temperature since its one-dimensional character.
Therefore we define a term ``quasi'' to refer low temperature regions
mainly dominated by ground state configuration. Hence $FR_{2}$ in
low temperature region is called as $qFR_{2}$, and so on.
As shown in~\cite{ph-bd,*Rojas2020}, pseudo-transitions occur for states 
near those phase boundaries whose residual entropy is a continuous function 
of the model parameters for at least one of the adjacent phases.
As discussed earlier, the state of the $FR_{2}-FM_{1}$ boundary coincides with the $FR_{2}$ state, 
so for the $qFR_{2}-qFM_{1}$ boundary, we get from eq.\eqref{eq:psd-eq} the following relation
\begin{equation}
{\rm e}^{\frac{K+h}{T_{p}}}={\rm e}^{-\frac{J}{T_{p}}}(q-2)+1,
\label{eq:GTp-fru}
\end{equation}
which we can simplify and write approximately in the form 
\begin{equation}
T_{p}=\frac{J+K+h}{\ln\left(q-2\right)}.\label{eq:LTp-fru}
\end{equation}
This is the known expression~\cite{Galisova,strk-cav,pseudo,ph-bd,*Rojas2020} 
for the pseudo-transition temperature:
\begin{equation}
E_{FM_{1}}-E_{FR_{2}}=T_{p}\left(\mathcal{S}_{FR_{2}}-\mathcal{S}_{FM_{1}}\right),
\end{equation}
where the energy and entropy per unit cell are given at zero temperature.
Since $E_{FM_{1}}=E_{FR_{2}}$ at the $qFR_{2}$-$qFM_{1}$ boundary,
$T_{p}$ tends to zero near to it.

Another phase boundary we focus is $qFM_{2}$-$qFM_{1}$. It is worth 
to mention that, the entropy of the $FM_{1}$ and $FM_{2}$ phases 
is zero, so the entropy is a continuous function for both adjacent phases.
For the $qFM_{2}$-$qFM_{1}$ boundary the eq.\eqref{eq:psd-eq} 
can be approximately written in the following form 
\begin{equation}
\frac{1}{T_{p}}\left(K+h\right)e^{J/T_{p}}=q-2.
\label{eq:Tp-unf}
\end{equation}

\begin{figure}
\includegraphics[scale=0.52]{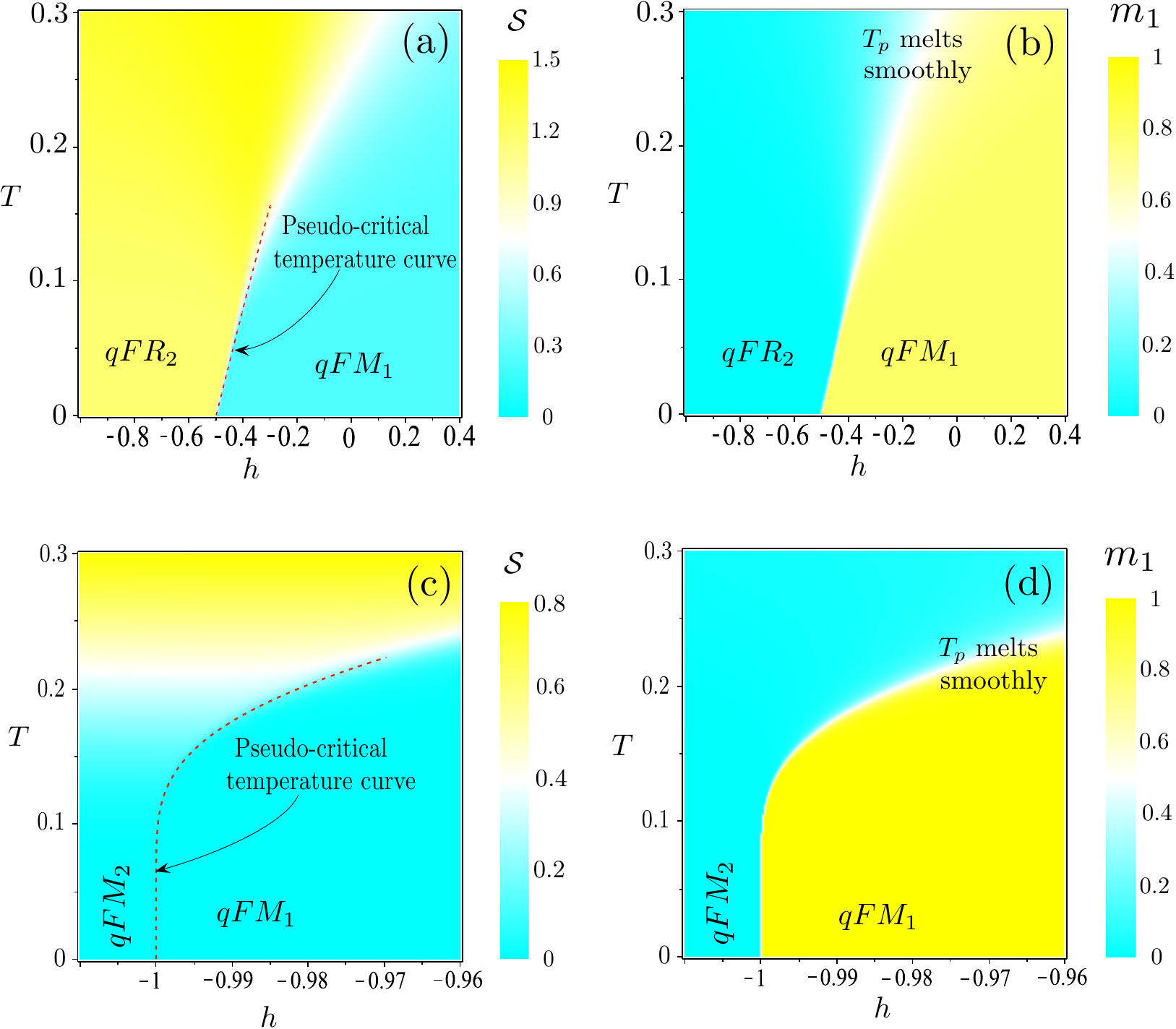}
\caption{\label{fig:Density-plot}Density plot: (a) for entropy in the plane
$T-h$, for $K=1$ and $J=-0.5$. (b) Magnetization $m_{1}$ for the
same set of parameters to the panel (a). In (c) entropy in the plane
$T-h$ for fixed $K=1$, $J=0.99$. In (d) magnetization for the
same set of parameter to the panel (c).}
\end{figure}

In Fig.~\ref{fig:Density-plot}a is reported the density plot of entropy
in the plane $T-h$, assuming fixed parameters $K=1$, $J=-0.5$.
Dashed curve describes the boundary $qFR_{2}$-$qFM_{1}$, which corresponds
to the pseudo-critical temperature $T_{p}$ as a function of $h$,
according eq.\eqref{eq:GTp-fru}. 
It can be seen that the curve is an almost straight line 
well represented by \eqref{eq:LTp-fru}. We can observe also
how the sharp boundary between quasi-phase melts smoothly
for higher temperature. Similar density plot is depicted in panel(b)
for the magnetization $m_{1}$ in the plane $T-h$ for the same set
of parameters in panel (a). Analogously, we analyze the phase boundary
between $qFM_{2}-qFM_{1}$ in panel (c), assuming fixed parameters
$K=1$, $J=0.99$ and $T=0.01$. The dashed line is given by eq.\eqref{eq:GTp-fru}
and nicely approximated by \eqref{eq:Tp-unf}, since there is no residual
entropy in the boundary the quasi-phase $qFM_{1}$ and $qFM_{2}$
leads to zero when temperature vanishes, by looking entropy we cannot
distinguishes the boundary of quasi-phase. However, in panel (d) we
illustrate the density plot of magnetization $m_{1}$, for the same
set of parameters to the panel (c), and we observe clearly a sharp
boundary between $qFM_{1}$ and $qFM_{2}$ regions, and this boundary
melts smoothly as soon as temperature increases.

\subsection{Entropy and Specific heat}

The entropy and the specific heat of the system can be obtained from the free energy~\eqref{eq:FE} by
\begin{equation}
\mathcal{S}=-\frac{\partial f}{\partial T}, \qquad C = -T \frac{\partial^2 f}{\partial T^2}.
\label{eq:s}
\end{equation}

Of particular interest is the behavior of thermodynamic characteristics 
near the pseudo-transition point. The general method for considering this issue 
was developed in Ref.~\cite{unv-cr-exp}. 
For the one-dimensional Potts model, it is possible to find 
an explicit form of approximation of the free energy 
and other thermodynamic quantities near $T_p$. 
Assuming that $\tau = (T-T_p)/T_p \ll 1$ and taking into account 
the equation~\eqref{eq:psd-eq}, we can write 
\begin{alignat}{1}
w_{1}= & \; e^{\beta(J+K+h)} = \tilde{w}_1^{\frac{1}{1+\tau}} 
\approx \tilde{w}_1 \left(1-\ln\tilde{w}_1 \, \tau\right) ,\label{eq:w1-appr}\\
w_{-1}= & \; q-2 + e^{\beta J} = q-2 + x_p^{\frac{1}{1+\tau}}
\approx \tilde{w}_1 - x_p \ln x_p \,\tau ,\label{eq:wm1-appr}
\end{alignat}
where 
\begin{equation}
\tilde{w}_{1} = \left.w_{1}\right|_{\tau=0} = q-2 + x_p , 
\end{equation}
and we define the parameters $x_p=e^{J/T_p}$ and $k_p=e^{K/T_p}$.
The value of $w_1 - w_{-1}$ is zero at $\tau=0$ due to equation~\eqref{eq:psd-eq}, 
so we get 
\begin{equation}
	w_1 - w_{-1} \approx -\tilde{w}_1 \, a \, \tau,
\end{equation}
where we introduce the parameter $a$ as
\begin{equation}
	a = \ln \tilde{w}_1 - \frac{x_p \ln x_p}{\tilde{w}_1} .
\end{equation}
Note that the parameter $a^{-1}$ define the slope of the pseudo-transition curve 
in the plane $T-h$ (see Fig.~\ref{fig:Density-plot}a,c), 
$dT_p/dh = a^{-1}$.

The condition $|w_1 - w_{-1}| \gg w_0$ which is met in the vicinity of $T_p$ causes 
quasi-singular behaviour of the derivative of square root in equations~(\ref{eq:Lmbd1},\ref{eq:Lmbd2}) 
at $\tau=0$. Assuming this is the case, we can write  the approximation
\begin{equation}
	4 w_0^2 = 4 (q-1) e^{\beta h} \approx 4 (q-1) \frac{\tilde{w}_1}{x_p k_p}.
	\label{eq:w0-appr}
\end{equation}
This allows us to write eigenvalues~(\ref{eq:Lmbd1},\ref{eq:Lmbd2}) near $\tau=0$ as 
\begin{equation}
	\lambda_{1,2} \approx \tilde{w}_1 
	\left[1 + \left(\tfrac{1}{2}a\tau-\ln\tilde{w}_1\right)\tau 
	\pm \tfrac{1}{2}\sqrt{a^2 \tau^2 + b^2} \right],
	\label{eq:lam12-appr}
\end{equation}
where $b^2 = 4 w_0^2 / \tilde{w}_1^2$,
and yields the approximation for the free energy~\eqref{eq:FE} in the vicinity of $T_p$: 
\begin{equation}
	f \approx -T_p \left[ \ln\tilde{w}_1 - \tfrac{1}{2}a\tau + \tfrac{1}{2}\sqrt{a^2 \tau^2 + b^2} \right].
	\label{eq:FE-appr}
\end{equation}

The expressions~(\ref{eq:lam12-appr},\ref{eq:FE-appr}) being exact at $\tau=0$ 
have a small deviation from rigorous expansions of $\lambda_{1,2}$ and $f$ in the tiny vicinity 
$0<|\tau|<\ln(x_p k_p)\,b^2/a^2\lll1$ due to the neglect of linear terms in equation~\eqref{eq:w0-appr}, 
but well approximate the functions $\lambda_{1,2}$ and $f$ at $\ln(x_p k_p)\,b^2/a^2<|\tau|\ll1/|\ln x_p|$. 
This implies a simple necessary condition for a pseudo-transition in the form $b/a\ll T_p/|J|$. 

For the entropy in the same region of $\tau$, using~\eqref{eq:s}, we obtain the following expression:
\begin{equation}
	\mathcal{S} \approx \frac{a}{2} \left(1 + \frac{a\tau}{\sqrt{a^2 \tau^2 + b^2}}\right), 
	\label{eq:S-appr}
\end{equation}
The equation~\eqref{eq:S-appr} describes the entropy jump in a small vicinity of $T_p$
\begin{equation}
	\Delta\mathcal{S}_p=
	\mathcal{S}\left(\tau>\tfrac{b}{a}\right)-\mathcal{S}\left(\tau<-\tfrac{b}{a}\right)=a,
	\label{eq:DSp}
\end{equation}
which may be related to the ``latent heat'' of pseudo-transition, 
$Q = T_p \Delta \mathcal{S}_p = a T_p$.

Second derivation of equation~\eqref{eq:FE-appr} by temperature gives 
the approximation of the specific heat near $T_p$,
\begin{equation}
	C \approx \frac{a^2 b^2}{2 \left(a^2 \tau^2 + b^2\right)^{\frac{3}{2}}}.
	\label{eq:C-appr}
\end{equation}
This allows us to estimate the maximum value of the specific heat in $T_p$ as 
\begin{equation}
	C_p = \tfrac{1}{2}a^2 b^{-1}.
\end{equation}
We can qualify the peak of the specific heat near $T_p$ by its 
half-width at half-maximum $\Psi_{\tau}$. 
From~\eqref{eq:C-appr} we find that $\Psi_{\tau}=\gamma b/a$, 
where $\gamma=\sqrt{2^{2/3}-1} \approx 0.7664$, 
and hence $\Psi_{\tau}\ll1$ due to a necessary condition for a pseudo-transition.

\begin{figure}
\includegraphics[width=0.23\textwidth]{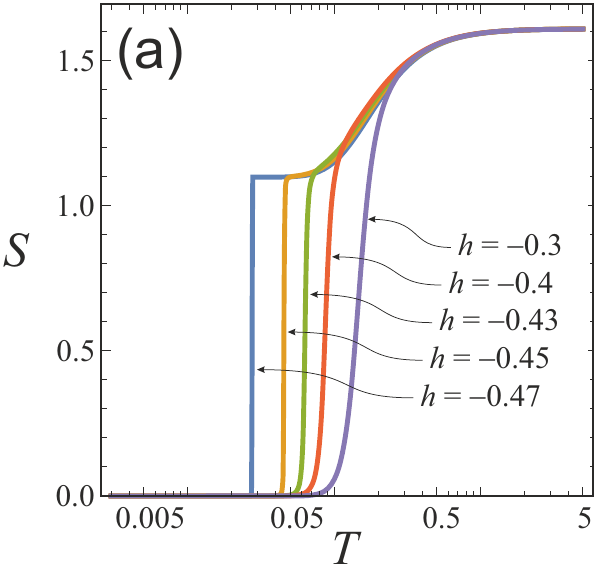} \includegraphics[width=0.23\textwidth]{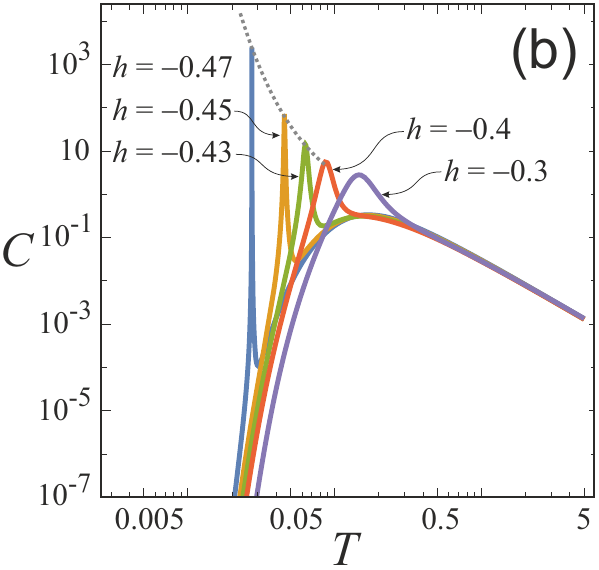} \\
\includegraphics[width=0.23\textwidth]{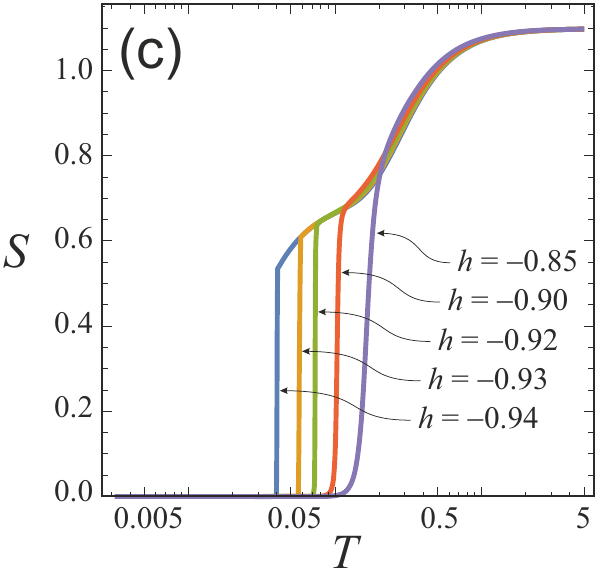} \includegraphics[width=0.23\textwidth]{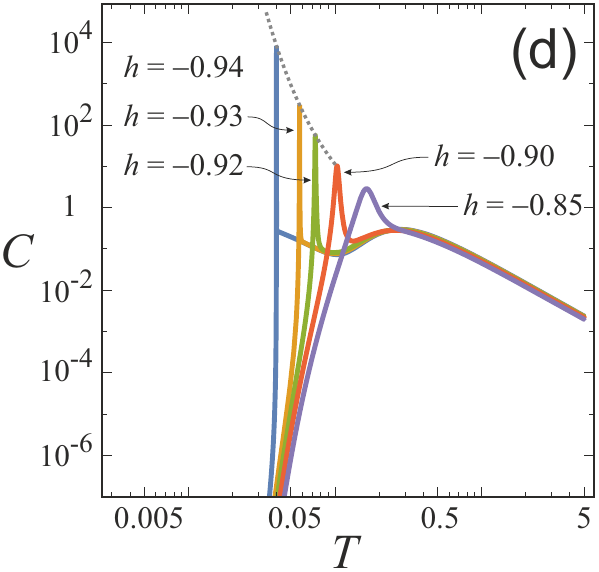} 
\caption{The entropy and specific heat of the $qFM_{1}$ states near the $FR_{2}$-$FM_{1}$ boundary
(a, b) with $q=5$, $J=-0.5$, $K=1$, and (c, d) with $q=3$, $J=-0.05$, $K=1$, 
in an external field $h$. 
The dotted lines in (b) and (d) show the magnitude of $C_p$ at $T=T_p$ given by 
equations~\eqref{eq:Cp1} and~\eqref{eq:Cp2} respectively.
}
\label{fig:scFRFM}
\end{figure}

The entropy and specific heat of the $qFM_{1}$ states 
having $q=5$, $J=-0.5$, $K=1$ in a given external field $h$ 
are shown in Fig.~\ref{fig:scFRFM}a,b. 
At sufficiently low temperatures, the $qFR_{2}$-$qFM_{1}$ pseudo-transition 
is observed, which is accompanied by a jump in entropy 
and a narrow peak in the specific heat.

For the $qFR_{2}$-$qFM_{1}$ transition at $q>3$, the condition 
$x_p \ll q-2$ is met, since $J<0$ and $|J|/T_p\gg1$, 
so approximately
\begin{equation}
	a = \ln(q-2), \qquad b=2\sqrt{\frac{q-1}{(q-2)x_p k_p}}.
\end{equation}
The entropy jump $\Delta\mathcal{S}_p=\ln(q-2)$ equals to the residual entropy 
of the $qFR_{2}$ phase (see Table~\ref{tab:mgs1}).
An expression for the maximum value of the specific heat, 
\begin{equation}
	C_p = \frac{\ln^2(q-2)}{4} \sqrt{\frac{q-2}{q-1}} \, e^{\frac{K-|J|}{2T_p}},
	\label{eq:Cp1}
\end{equation}
shows that $C_p$ decreases for the states close to the point $P_1$ in Fig.~\ref{fig:PDchain1}c. 
The magnitudes of the specific heat peaks~\eqref{eq:Cp1} 
are shown in Fig.~\ref{fig:scFRFM}b with dotted lines.
It is interesting to note, that 
$C_{p}\Psi_{\tau}=\tfrac{1}{2}\gamma\ln(q-2)={\rm const}(T_p)$, 
so, having a finite height, the specific heat peak tends to the delta function
near the $FR_{2}$-$FM_{1}$ boundary with the $T_p$ lowering.

In Fig.~\ref{fig:scFRFM}c,d the entropy and specific heat of the $qFM_{1}$ states 
having $q=3$, $J=-0.05$, $K=1$ in a given external field $h$ are shown. 
The case $q=3$ is special, since 
the residual entropy of the $FR_{2}$-$FM_{1}$ boundary 
is a continuous function for both adjacent phases. 
For the parameter $a$ we obtain
\begin{equation}
	a = \left(1-\ln x_p\right) x_p, 
	\label{eq:a2}
\end{equation}
so the necessary condition for a pseudo-transition 
$b/a \propto e^{(3|J|-K)/2T_p} \ll T_p/|J|$ is met if $K>3|J|$. 
Since in this case $J<0$, the entropy jump~\eqref{eq:DSp} drops 
with decreasing of $T_p$ due to the exponent in $x_p=e^{J/T_p}$, 
as can be seen from Fig.~\ref{fig:scFRFM}c. 
A maximal value of the specific heat in $T_p$ approximately has the following form
\begin{equation}
	C_p = \frac{\left(1-J/T_p\right)^2}{4\sqrt{2}} \, e^{\frac{K-5|J|}{2T_p}}.
	\label{eq:Cp2}
\end{equation}
From~\eqref{eq:Cp2} we may conclude that if $K>5|J|$ 
the pseudo-transition is accompanied with exponentially high peak of the specific heat. 
From a general point of view, this issue was discussed in detail in~\cite{ph-bd,*Rojas2020}.

\begin{figure}
\includegraphics[width=0.23\textwidth]{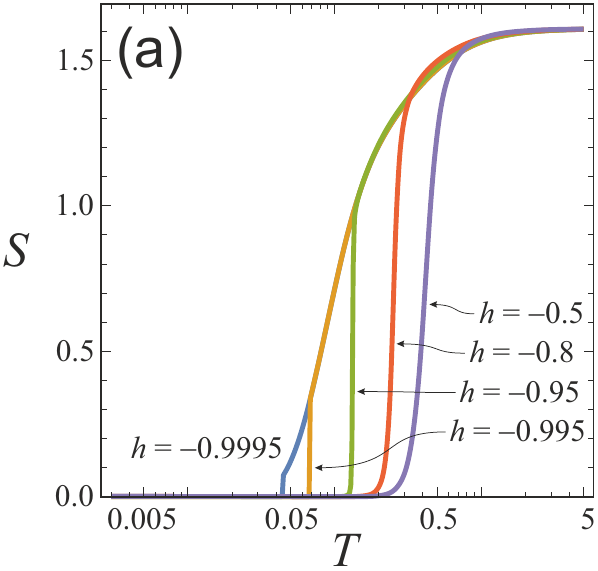} \includegraphics[width=0.23\textwidth]{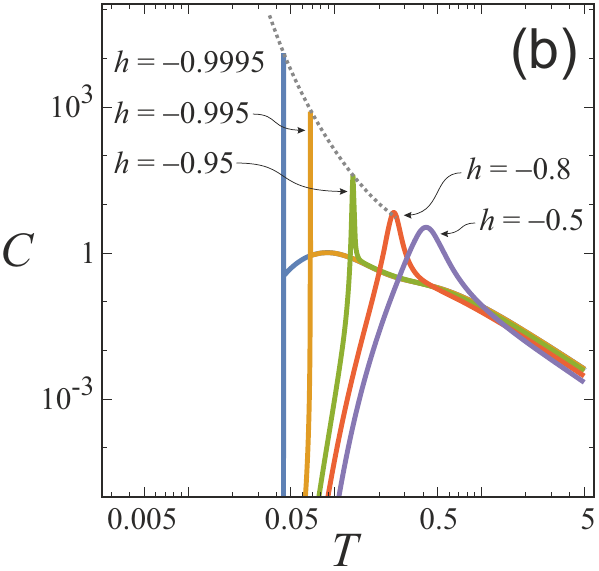} \\
\includegraphics[width=0.23\textwidth]{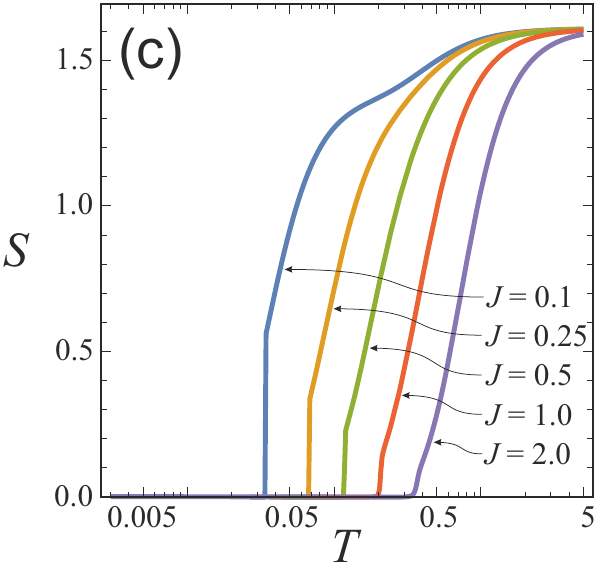} \includegraphics[width=0.23\textwidth]{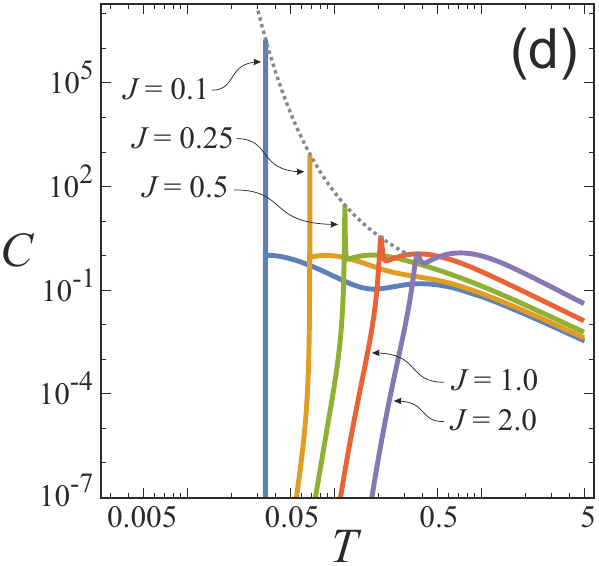} 
\caption{The entropy and specific heat of the $qFM_{1}$ states near the $FM_{1}$-$FM_{2}$ boundary
(a, b) with $q=5$, $J=0.25$, $K=1$ in an external field $h$,
and (c, d) with $q=5$, $h=-0.995$, $K=1$ for the different values of the coupling parameter $J$.
The dotted lines in (b) and (d) show the magnitude of $C_p$ at $T=T_p$ given by 
equation~\eqref{eq:Cp3}.
}
\label{fig:scFMM}
\end{figure}

Figure~\ref{fig:scFMM} shows the temperature dependencies 
of the entropy and specific heat for states having parameters 
close to the $FM_{1}$-$FM_{2}$ boundary for $K=1$. 
One set of states, shown in Fig.~\ref{fig:scFMM}a,b, 
has $J=0.25$ and different values of the external field $h$, 
and the states in another set, shown in Fig.~\ref{fig:scFMM}c,d, 
have the same $h$ and differ in the coupling constant $J$.

Near the $FM_{1}$-$FM_{2}$ boundary, 
for which the residual entropy is zero, as for both adjacent phases, 
we get
\begin{equation}
	a = (q-2)\frac{1+\ln x_p}{x_p}, \qquad b=\frac{2}{x_p}\sqrt{\frac{q-1}{k_p}}.
	\label{eq:a3}
\end{equation}
The condition for a pseudo-transition $b/a \propto e^{-K/2T_p} \ll T_p/|J|$ will met only if $K>0$. 
In this case $J>0$, so the entropy jump $\Delta \mathcal{S}_p = a \propto e^{-J/T_p}$ 
decreases exponentially with decreasing $T_p$. 
This effect is shown in Fig.~\ref{fig:scFMM}a. 
The specific heat in $T_p$ approximately is given by
\begin{equation}
	C_p = \frac{(q-2)^2\left(1+J/T_p\right)^2}{4\sqrt{q-1}} \, e^{\frac{K-2J}{2T_p}}.
	\label{eq:Cp3}
\end{equation}
Equation~\eqref{eq:Cp3} shows that for $K>2J$ the $qFM_{2}$-$qFM_{1}$ 
pseudo-transition entails an exponentially high peak of the specific heat 
if the pseudo-transition temperature is small enough.
The decrease in the peak of the specific heat near $T_p$ with increasing $J$ 
is shown in Fig.~\ref{fig:scFMM}d.

\subsection{Magnetization and Magnetic susceptibility}

It is important to study the magnetization property of the present
model. In order to obtain a general average $\langle\delta_{\sigma,\mu}\rangle$,
let us define the following operator

\begin{equation}
\boldsymbol{m}_{\mu}=\sum_{j=1}^{q}\delta_{j,\mu}|j\rangle\langle j|=|\mu\rangle\langle\mu|.\label{eq:-op-m_mu}
\end{equation}

Therefore, the average of $\boldsymbol{m}_{\mu}$ becomes
\begin{equation}
\langle\delta_{\sigma,\mu}\rangle=m_{\mu}=Z_{N}^{-1}\mathop{\mathrm{tr}}\left(\boldsymbol{m}_{\mu}\mathbf{V}^{N}\right).\label{eq:<delta>}
\end{equation}

Expressing \eqref{eq:-op-m_mu}, in eigenstate basis given in (\ref{eq:u1}-\ref{eq:u3}),
we have

\begin{equation}
\tilde{\boldsymbol{m}}_{\mu}=\sum_{j_{1},j_{2}=1}^{q}\langle u_{j_{1}}|\boldsymbol{m}_{\mu}|u_{j_{2}}\rangle|u_{j_{1}}\rangle\langle u_{j_{2}}|,\label{eq:Ptm_mu}
\end{equation}
with

\begin{equation}
\langle u_{j_{1}}|\boldsymbol{m}_{\mu}|u_{j_{2}}\rangle=\langle u_{j_{1}}|\mu\rangle\langle\mu|u_{j_{2}}\rangle,
\end{equation}
and denoting $\langle\mu|u_{j}\rangle=c_{\mu,j}$, we can write \eqref{eq:Ptm_mu}
as 

\begin{equation}
\tilde{\boldsymbol{m}}_{\mu}=\sum_{j_{1},j_{2}=1}^{q}c_{\mu,j_{1}}^{*}c_{\mu,j_{2}}|u_{j_{1}}\rangle\langle u_{j_{2}}|.\label{eq:tm_mu}
\end{equation}
In a similar way, the transfer matrix in eigenstate basis (\ref{eq:u1}-\ref{eq:u3}),
becomes

\begin{equation}
\tilde{\mathbf{V}}=\sum_{j=1}^{q}\lambda_{j}|u_{j}\rangle\langle u_{j}|.\label{eq:tV}
\end{equation}

Hence, substituting \eqref{eq:tm_mu} and \eqref{eq:tV} in eq.\eqref{eq:<delta>},
we obtain 
\begin{equation}
m_{\mu}=\frac{1}{Z_{N}}\!\sum_{j_{1}=1}^{q}\langle u_{j_{1}}|\negmedspace\sum_{j_{2},j_{3}=1}^{q}\negmedspace c_{\mu,j_{2}}^{*}c_{\mu,j_{3}}|u_{j_{2}}\rangle\langle u_{j_{3}}|\lambda_{j_{3}}^{N}|u_{j_{1}}\rangle.\label{eq:<tm_mu>}
\end{equation}
In thermodynamic limit ($N\rightarrow\infty$) we can simplify \eqref{eq:<tm_mu>}.
Therefore, writing the magnetization $m_{\mu}$ in terms of $c_{\mu,j}$,
it reduces to

\begin{equation}
m_{\mu}=c_{\mu,1}^{*}c_{\mu,1}=|c_{\mu,1}|^{2}.\label{eq:m_mu}
\end{equation}
The explicit expression of coefficients are given by

\begin{alignat}{1}
c_{1,1}=\cos\phi,\quad & c_{\mu,1}=\frac{\sin\phi}{\sqrt{q-1}},\label{eq:cmu1}\\
c_{1,2}=\sin\phi,\quad & c_{\mu,2}=\frac{\cos\phi}{\sqrt{q-1}},\label{eq:cmu2}
\end{alignat}
where $\mu=\{2,\dots,q\}$. 

Analogously the remaining coefficients are written by

\begin{alignat}{1}
c_{\mu,j}= & \begin{cases}
0; & \mu=\{1,j+1,\dots,q\}\\
\frac{1}{\sqrt{(j-1)(j-2)}}; & \mu=\{2,\dots,j-1\}\\
\sqrt{\frac{j-1}{j-2}}; & \mu=j
\end{cases},\label{eq:cmuj}
\end{alignat}
where we consider $j=\{3,\cdots,q\}$. 

As a consequence, we can get the magnetization by using \eqref{eq:cmu1}
in \eqref{eq:m_mu}, which becomes
\begin{alignat}{1}
m_{1}=\cos^{2}\phi,\quad & m_{\mu}=\frac{\sin^{2}\phi}{q-1}.\label{eq:mu1}
\end{alignat}

Additionally, from above result, we get the following identity 
\begin{equation}
m_{1}+(q-1)m_{\mu}=1,\quad\mu=\{2,\dots,q\}.\label{eq:ident-m1-m_mu}
\end{equation}

Alternatively one can obtain $m_{1}$ taking the derivative of free
energy with respect to external field $h$, 
\begin{eqnarray}
m_{1} & = & \left\langle \delta_{\sigma,1}\right\rangle =-\frac{\partial f}{\partial h},\label{eq:m}
\end{eqnarray}
and $m_{\mu}$ we can obtain from \eqref{eq:ident-m1-m_mu}.

On the other hand, the magnetic susceptibility $\chi_{1}$ can be
obtained deriving \eqref{eq:m}, which results in 

\begin{equation}
\chi_{1}=\frac{1}{4T}\sin(2\phi)^{3}\left(\frac{d_{1}+d_{2}+(q-2)t_{2}}{2t_{1}\sqrt{q-1}}\right),
\end{equation}
Similarly, one can obtain for $\mu=\{2,\dots,q\}$,

\begin{equation}
\chi_{\mu}=\frac{\partial m_{\mu}}{\partial h},
\end{equation}
and we have the following relation
\begin{equation}
\chi_{\mu}=-\frac{1}{q-2}\chi_{1},
\end{equation}
for $\mu=\{2,\dots,q\}$.

To find an approximate expression for the magnetization near $T_p$, 
we can write it by using equations~\eqref{eq:FE} and~\eqref{eq:m} in the form
\begin{equation}
	m_1 = \frac{T}{\lambda_1} \frac{\partial \lambda_1}{\partial h}.
	\label{eq:m1-1}
\end{equation}
When calculating the derivative with respect to $h$, we take into account that
$\tfrac{\partial}{\partial h} w_{1}=\beta w_1$, $\tfrac{\partial}{\partial h} w_{-1} = 0$, 
$\tfrac{\partial}{\partial h} w_{0}^2 = \beta w_{0}^2$, 
and obtain
\begin{equation}
	m_1 = \frac{1}{2\lambda_1} \left(w_1 + \frac{(w_1-w_{-1})w_1+2w_0^2}{(w_1-w_{-1})^2+4w_0^2}\right).
\end{equation}
Using equations~(\ref{eq:w1-appr}-\ref{eq:w0-appr}) 
and leaving only the leading terms, we find an approximation for $m_1$ 
in the following form:
\begin{equation}
	m_1 \approx \frac{1}{2} \left(1 - \frac{a\tau}{\sqrt{a^2 \tau^2 + b^2}}\right).
	\label{eq:m1-appr}
\end{equation}
Equation~\eqref{eq:m1-appr} describes the jump in magnetization 
from $m_1=0$ at $\tau>b/a$ to $m_1=1$ at $\tau<-b/a$ 
in the small vicinity of $T_p$ for both $qFR_{2}$-$qFM_{1}$ and 
$qFM_{2}$-$qFM_{1}$ pseudo-transitions.

Similarly, for the susceptibility $\chi_1$ we may write, 
differentiating~\eqref{eq:m1-1} with respect to $h$:
\begin{equation}
	\chi_1 = -\frac{m_1^2}{T} + \frac{T}{\lambda_1} \frac{\partial^2 \lambda_1}{\partial h^2}.
	\label{eq:chi1-1}
\end{equation}
The quasi-singular behavior in $T_p$ is caused by the second term in~\eqref{eq:chi1-1}.
Using the same steps as in deriving~\eqref{eq:m1-appr} 
and leaving the main contributions, we come to the following approximations  
for the susceptibility near $T_p$:
\begin{equation}
	\chi_1 \approx \frac{b^2}{2T_p \left(a^2 \tau^2 + b^2\right)^{\frac{3}{2}}},
	\label{eq:chi1-appr}
\end{equation}
and its maximum value:
\begin{equation}
	\chi_{1,p} = \frac{b^{-1}}{2T_p}.
	\label{eq:chi1p}
\end{equation}

\begin{figure}
\includegraphics[width=0.23\textwidth]{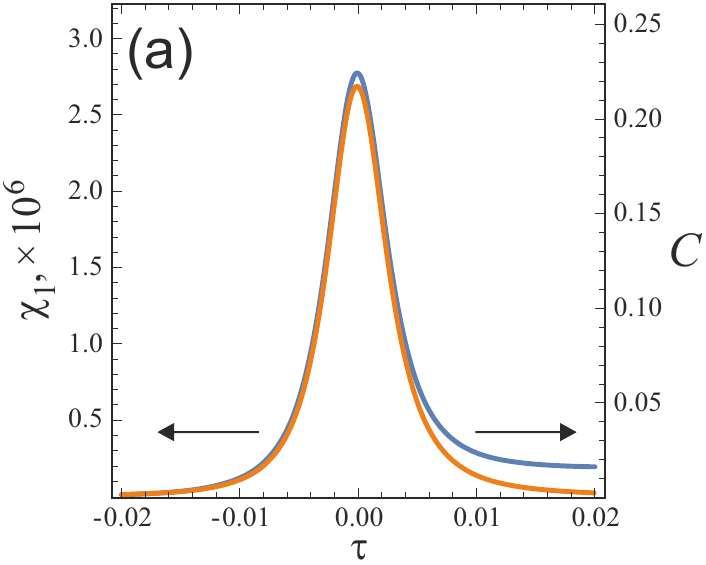} \  \includegraphics[width=0.23\textwidth]{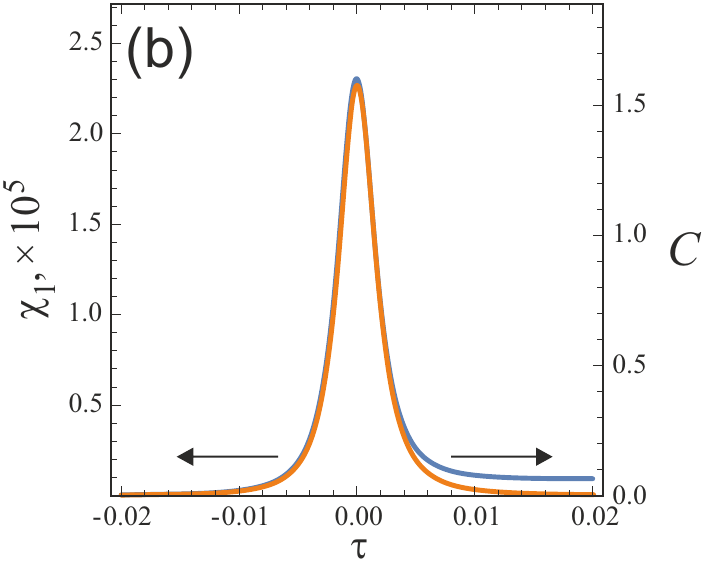} \\
\caption{The specific heat (blue line)  and susceptibility (orange line) peaks 
in the vicinity of pseudo-transition points at
(a) $q=3$, $J=-0.25$, $K=1$, $h=-0.749995$,
and (b) $q=5$, $J=1$, $K=1$, $h=-0.9999$.
}
\label{fig:sim}
\end{figure}

Comparing~\eqref{eq:C-appr} and~\eqref{eq:chi1-appr}, we see that dependencies 
of the specific heat and susceptibility in the vicinity of $T_p$
are similar: they have the same half-width at half-maximum $\Psi_{\tau}$ 
and differ by a scale factor $\alpha$, which is the ratio of the specific heat and susceptibility 
in the pseudo-transition point 
\begin{equation}
	\alpha = \frac{C_p}{\chi_{1,p}} = a^2 T_p.
\end{equation}
If for the $qFR_{2}$-$qFM_{1}$ pseudo-transition at $q>3$ 
the factor $\alpha$ decreases linearly with lowering $T_p$, 
for the $qFR_{2}$-$qFM_{1}$ pseudo-transition at $q=3$ and the $qFM_{2}$-$qFM_{1}$ pseudo-transition, 
it decreases exponentially due to dependence of $a$.
In both the latter cases, for some parameters, 
the giant magnitude of $\chi_{1,p}$ can be realized at the low magnitude of $C_p$.
Indeed, for the $qFR_{2}$-$qFM_{1}$ pseudo-transition at $q=3$, we have 
$\chi_{1,p} \propto e^{(K-|J|)/2T_p}$, so the giant peak of susceptibility 
exists at sufficiently low $T_p$ in the entire pseudo-transition region  
defined, as it was found earlier (see equation~\eqref{eq:a2}),  
by the condition $K>3|J|$, but for the specific heat it becomes 
exponentially high only if $K>5|J|$ due to equation~\eqref{eq:Cp2}.
This case is shown in Fig.~\ref{fig:sim}a, 
where in the vicinity of the $qFR_{2}$-$qFM_{1}$ pseudo-transition 
at $q=3$, $J=-0.25$, $K=1$,  
the peaks of the specific heat and susceptibility differ in amplitude 
by seven orders of magnitude and coincide in shape.
In turn, for the $qFM_{2}$-$qFM_{1}$ pseudo-transition we have 
$\chi_{1,p} \propto e^{(K+2J)/2T_p}$. 
Taking into account~\eqref{eq:a3}, we can conclude that 
the giant peak of susceptibility exists for all $J>0$ at $K>0$, 
while for the specific heat, the giant peak  exists only at $0<2J<K$, 
as it follows from equation~\eqref{eq:Cp3}.
The similarity of the specific heat and susceptibility peaks 
in the vicinity of the $qFM_{2}$-$qFM_{1}$ pseudo-transition 
at $q=5$, $J=1$, $K=1$, is shown in Fig.~\ref{fig:sim}b.

\begin{figure}
\includegraphics[width=0.23\textwidth]{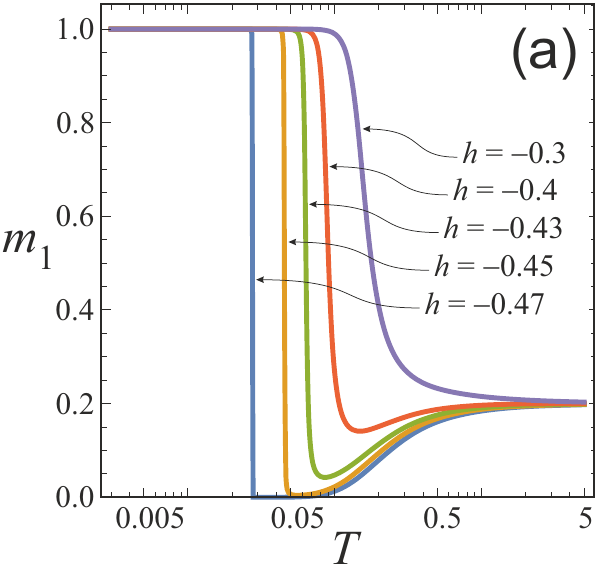} \includegraphics[width=0.23\textwidth]{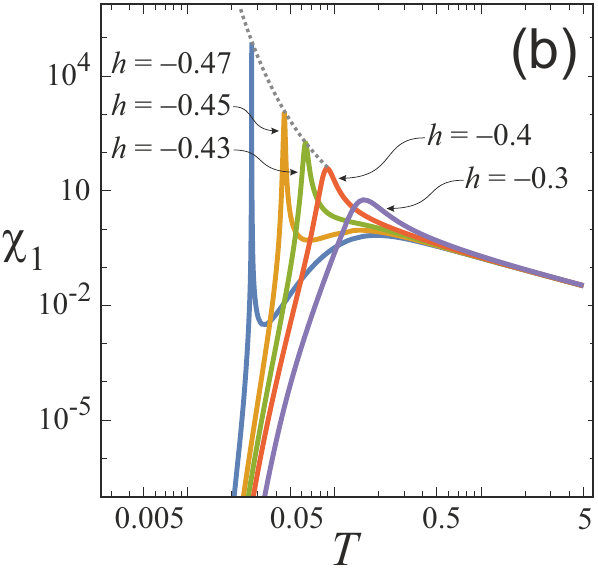} \\
\includegraphics[width=0.23\textwidth]{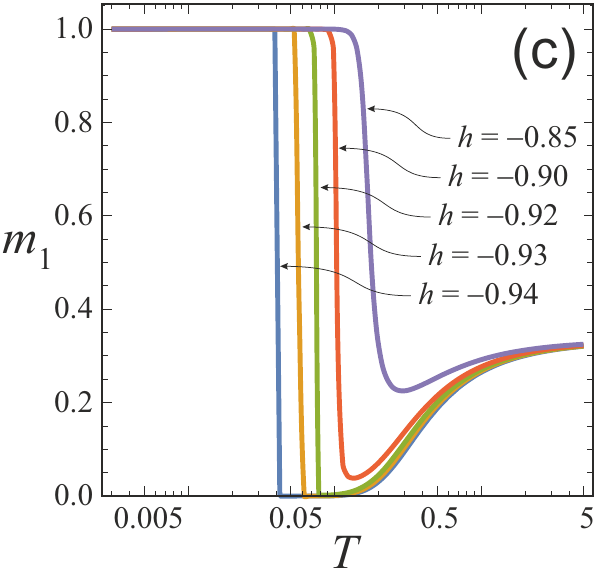} \includegraphics[width=0.23\textwidth]{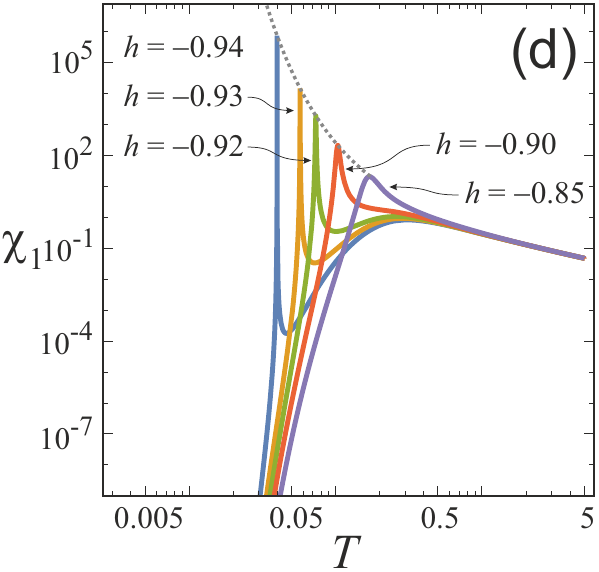} 
\caption{The magnetic moment and susceptibility of the $qFM_{1}$ states near the $FR_{2}$-$FM_{1}$ boundary
(a, b) with $q=5$, $J=-0.5$, $K=1$, and (c, d) with $q=3$, $J=-0.05$, $K=1$, 
in an external field $h$. 
The dotted lines in (b) and (d) show the magnitude of $\chi_{1,p}$ at $T=T_p$ given by 
equation~\eqref{eq:chi1p}.
}
\label{fig:mchiFRFM}
\end{figure}

\begin{figure}
\includegraphics[width=0.23\textwidth]{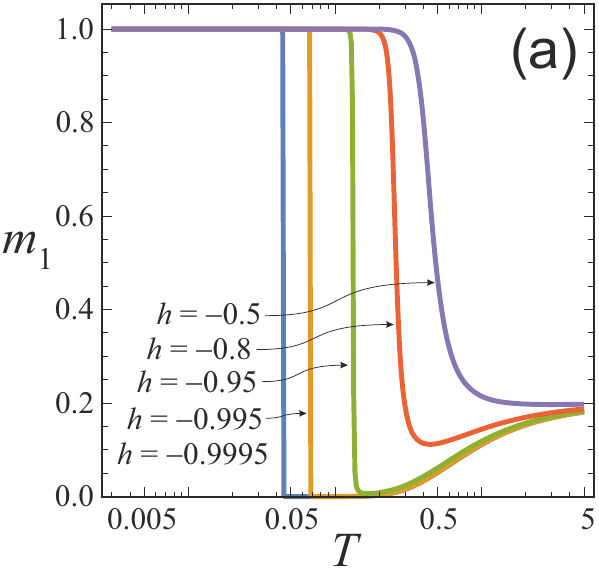} \includegraphics[width=0.23\textwidth]{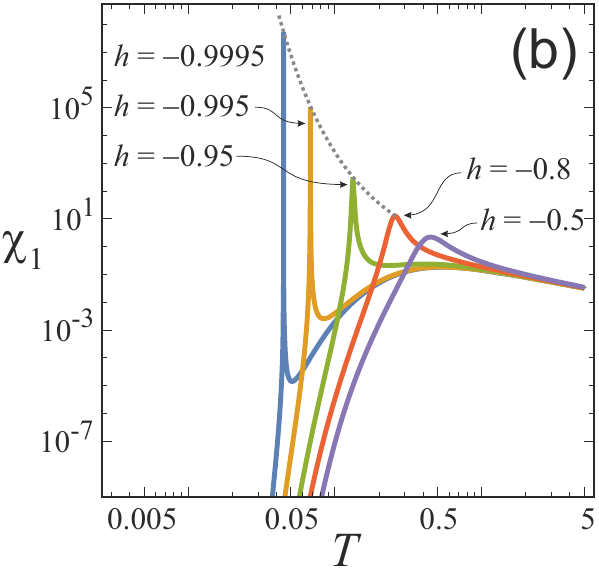} \\
\includegraphics[width=0.23\textwidth]{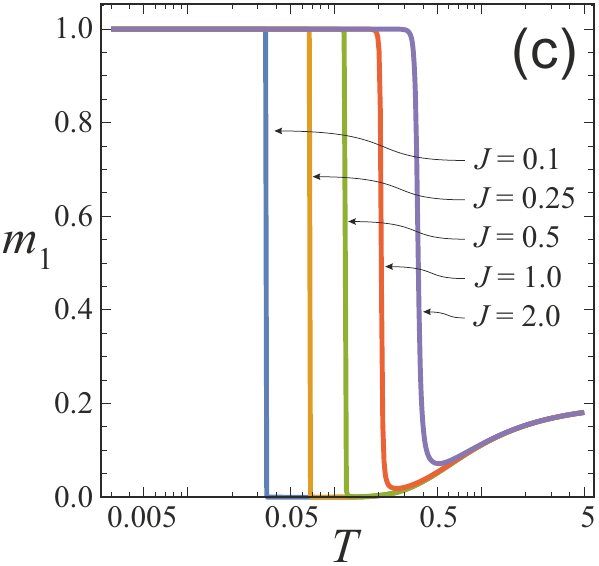} \includegraphics[width=0.23\textwidth]{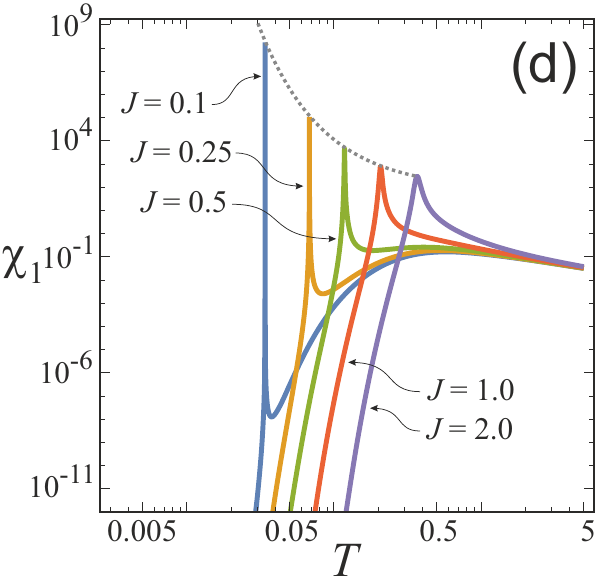}
\caption{The magnetic moment and susceptibility of the $qFM_{1}$ states near the $FM_{1}$-$FM_{2}$ boundary
(a, b) with $q=5$, $J=0.25$, $K=1$ in an external field $h$,
and (c, d) with $q=5$, $h=-0.995$, $K=1$ for the different values of the coupling parameter $J$.
The dotted lines in (b) and (d) show the magnitude of $\chi_{1,p}$ at $T=T_p$ given by 
equation~\eqref{eq:chi1p}.
}
\label{fig:mchiFMM}
\end{figure}

Temperature dependencies of the magnetic moment and susceptibility 
of the $qFM_{1}$ states near the $FR_{2}$-$FM_{1}$ boundary
are shown in Fig.~\ref{fig:mchiFRFM}. 
These states both for $q=5$, $J=-0.5$ and $q=3$, $J=-0.05$ exhibit 
the $qFR_{2}$-$qFM_{1}$ pseudo-transition with the continuous jump 
in the magnetic moment and the exponentially high peak of susceptibility. 
Figure~\ref{fig:mchiFMM} shows the magnetic moment 
and susceptibility for the same set of states as in Fig.~\ref{fig:scFMM}.
The dotted lines in Fig.~\ref{fig:mchiFRFM}b,d and in Fig.~\ref{fig:mchiFMM}b,d 
show an approximate value of susceptibility in the pseudo-transition point 
defined by equation~\eqref{eq:chi1p}.

\begin{figure}
 \includegraphics[width=0.23\textwidth]{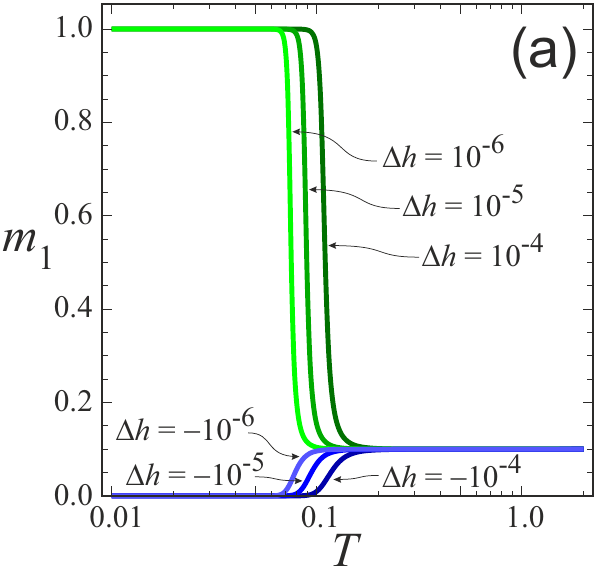} \includegraphics[width=0.23\textwidth]{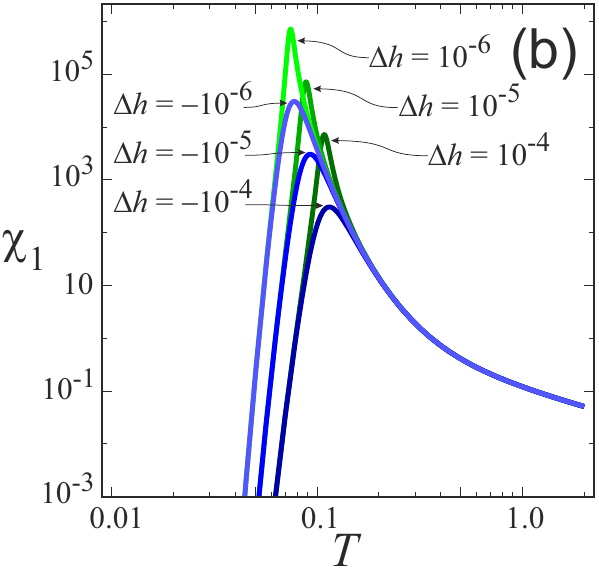} \\
 \includegraphics[width=0.23\textwidth]{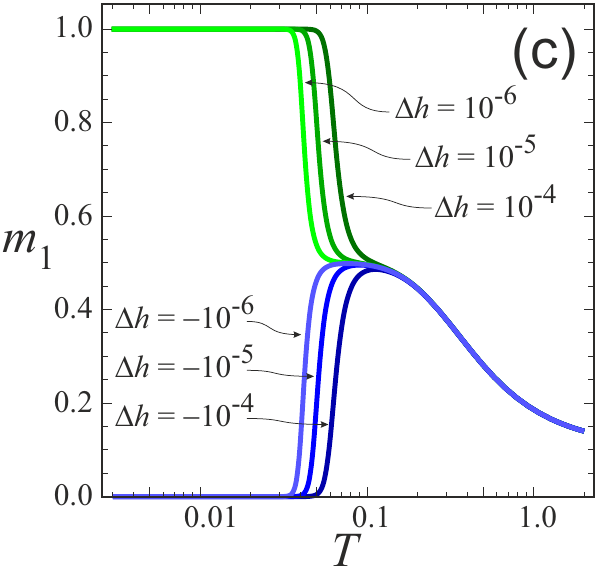} \includegraphics[width=0.23\textwidth]{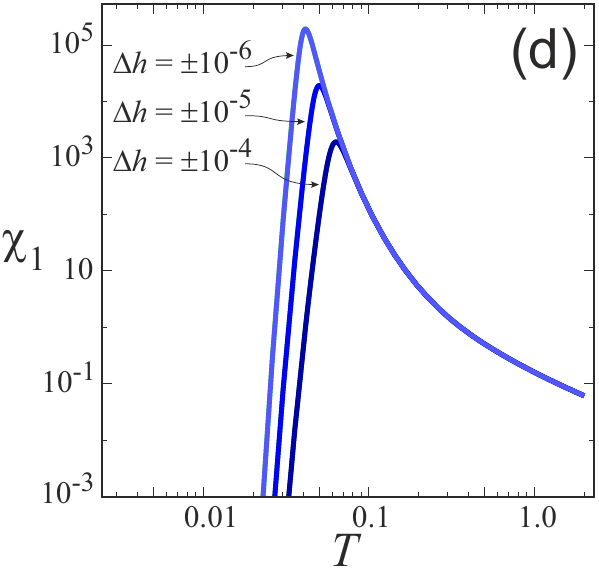}
\caption{The magnetic moment and susceptibility near the $FM_{1}$-$FM_{2}$
boundary for $qFM_{1}$ and $qFM_{2}$ states (green and blue lines)  with $q=10$, $J=1$, 
(a, b) for $K=0$, in an external field equal to $h=\Delta h$,
and (c, d) for $K=-1$, in an external field equal to $h = 1+\Delta h$. }
\label{fig:mchiFM1FM2}
\end{figure}

Temperature dependencies of the magnetic moment, and susceptibility
for states near the $FM_{1}$-$FM_{2}$ boundary at $K=0$ and $K<0$
are shown in Fig.~\ref{fig:mchiFM1FM2}.
It can be seen the qualitative difference from the case $K>0$.
At $K\le0$, the magnetic moment changes with decreasing temperature from 
the value at the $FM_{1}$-$FM_{2}$ boundary, 
$m_1=\tfrac{1}{q-1}$ at $K=0$ or $m_1=\tfrac{1}{2}$ at $K<0$, 
to the values of the magnetic moment in the $FM_{1}$ or $FM_{2}$ phase, 
which are equal to $1$ and $0$. 
Because of this, the susceptibility peak is observed in both the $qFM_{1}$ 
and $qFM_{2}$ states.

As shown if Fig.~\ref{fig:mchiFM1FM2}b,d, 
the peak of susceptibility at $K\le0$ becomes arbitrarily
large in height for the $qFM_{1}$ and $qFM_{2}$ states, which are
quite close to the $FM_{1}$-$FM_{2}$ boundary, but it is significantly
wider compared to the case of a pseudo-transition at $K>0$, or near the
$qFR_{2}$-$qFM_{1}$ pseudo-transition. 
The extremely narrow peak, which is characteristic of a pseudo-transition,
exists only for $K>0$. 
This follows formally from equation~\eqref{eq:a3} and the necessary inequality $b/a\ll T_p/|J|$.
The physical reason for this is the phase separation 
in the $FM_{1}$-$FM_{2}$ boundary at $K\le0$, which causes an intermediate 
value of the magnetic moment at the $FM_{1}$-$FM_{2}$ boundary and 
its jump in both the $qFM_{1}$ and $qFM_{2}$ states. 
The one-dimensional ferromagnetic Ising model has the same properties: 
zero magnetization of the ground state in the absence of an external field 
is achieved by splitting into ferromagnetic domains with opposite magnetization, 
and this state represents the boundary on the phase diagram 
between phases with magnetizations equal $+1$ and $-1$ depending on the sign 
of the external field. 
This situation can be considered as intermediate between microscopic mixing 
of neighboring phases at the phase boundary when there is no pseudo-transition, 
and a pure phase equal to one of the neighboring phases 
when the pseudo-transition is realized.

\section{\label{sec:PDF}Pair distribution correlation function}

In order to accomplish our analysis, we consider the pair distribution
correlation function, which is defined as follows

\begin{equation}
\Gamma_{\mu,\mu'}^{(r)}=\langle\delta_{\sigma_{i},\mu}\delta_{\sigma_{i+r},\mu'}\rangle-\langle\delta_{\sigma_{i},\mu}\rangle\langle\delta_{\sigma_{i+r},\mu'}\rangle,
\end{equation}
where $\Gamma_{\mu,\mu'}^{(r)}\equiv\Gamma_{\mu,\mu'}(\sigma_{i},\sigma_{i+r})$. 

If the system is translationally invariant, we have $\langle\delta_{\sigma_{i},\mu}\rangle=\langle\delta_{\sigma_{i+r},\mu}\rangle=\langle\delta_{\sigma,\mu}\rangle=m_{\mu}$,
and $\Gamma_{\mu,\mu'}^{(r)}$ depends only on the distance $(r)$.

Now let us defined the average of PDF,

\begin{equation}
g_{\mu,\mu'}^{(r)}=\langle\delta_{\sigma_{i},\mu}\delta_{\sigma_{i+r},\mu'}\rangle=\langle\boldsymbol{m}_{\mu}\boldsymbol{m}_{\mu'}\rangle,\label{eq:pdf}
\end{equation}
so, in natural basis is expressed by 
\begin{equation}
g_{\mu,\mu'}^{(r)}=Z_{N}^{-1}\mathop{\mathrm{tr}}\left(\boldsymbol{m}_{\mu}\mathbf{V}^{r}\boldsymbol{m}_{\mu'}\mathbf{V}^{N-r}\right)
\end{equation}
whereas in eigenstate basis becomes

\begin{equation}
g_{\mu,\mu'}^{(r)}=Z_{N}^{-1}\mathop{\mathrm{tr}}\left(\tilde{\boldsymbol{m}}_{\nu}\mathbf{\tilde{V}}^{r}\tilde{\boldsymbol{m}}_{\tau}\tilde{\mathbf{V}}^{N-r}\right).
\end{equation}
Writing in terms of eigenstates 
\begin{alignat}{1}
g_{\mu,\mu'}^{(r)}= & Z_{N}^{-1}\sum_{j_{1}=1}^{q}\langle u_{j_{1}}|\sum_{j_{2},j_{3}=1}^{q}c_{\mu,j_{2}}^{*}c_{\mu,j_{3}}|u_{j_{2}}\rangle\langle u_{j_{3}}|\lambda_{j_{3}}^{r}\nonumber \\
 & \times\sum_{j_{4},j_{5}=1}^{q}c_{\mu',j_{4}}^{*}c_{\mu',j_{5}}|u_{j_{4}}\rangle\langle u_{j_{5}}|\lambda_{j_{5}}^{N-r}|u_{j_{1}}\rangle\label{eq:g_coeff}
\end{alignat}
and simplifying \eqref{eq:g_coeff}, we have

\begin{equation}
g_{\mu,\mu'}^{(r)}=Z_{N}^{-1}\sum_{j_{1}=1}^{q}\sum_{j_{2}=1}^{q}c_{\mu,j_{1}}^{*}c_{\mu,j_{2}}\lambda_{j_{2}}^{r}c_{\mu',j_{2}}^{*}c_{\mu',j_{1}}\lambda_{j_{1}}^{N-r},
\end{equation}
in thermodynamic limit $(Z_{N}\rightarrow\lambda_{1}^{N})$. The above
expression reduces to

\begin{equation}
g_{\mu,\mu'}^{(r)}=\sum_{j_{1}=1}^{q}\sum_{j_{2}=1}^{q}c_{\mu,j_{1}}^{*}c_{\mu,j_{2}}\Bigl(\tfrac{\lambda_{j_{2}}}{\lambda_{1}}\Bigr)^{r}c_{\mu',j_{2}}^{*}c_{\mu',j_{1}}\left(\tfrac{\lambda_{j_{1}}}{\lambda_{1}}\right)^{N-r},
\end{equation}
manipulating conveniently, we have

\begin{equation}
g_{\mu,\mu'}^{(r)}=c_{\mu,1}^{*}c_{\mu,1}c_{\mu',1}^{*}c_{\mu',1}+\sum_{j=2}^{q}c_{\mu,1}^{*}c_{\mu,j}\Bigl(\tfrac{\lambda_{j}}{\lambda_{1}}\Bigr)^{r}c_{\mu',j}^{*}c_{\mu',1},
\end{equation}
or even in terms of eq.\eqref{eq:<delta>}, we have

\begin{equation}
g_{\mu,\mu'}^{(r)}=m_{\mu}m_{\mu'}+c_{\mu,1}^{*}c_{\mu',1}\sum_{j=2}^{q}c_{\mu,j}c_{\mu',j}^{*}\Bigl(\tfrac{\lambda_{j}}{\lambda_{1}}\Bigr)^{r}.
\end{equation}

Therefore, we can write the correlation function as follows

\begin{multline}
\Gamma_{\mu,\mu'}^{(r)}=c_{\mu,1}^{*}c_{\mu',1}\sum_{j=2}^{q}c_{\mu,j}c_{\mu',j}^{*}\Bigl(\tfrac{\lambda_{j}}{\lambda_{1}}\Bigr)^{r}=\\
=c_{\mu,1}^{*}c_{\mu',1}\Bigl\{ c_{\mu,2}c_{\mu',2}^{*}\left(\tfrac{\lambda_{2}}{\lambda_{1}}\right)^{r}+\left(\tfrac{\lambda_{3}}{\lambda_{1}}\right)^{r}\sum_{j=3}^{q}c_{\mu,j}c_{\mu',j}^{*}\Bigr\}.\label{eq:rGmm}
\end{multline}
Note that for $q=2$, the last term in \eqref{eq:rGmm} ceases to
exist.

Taking into account the orthogonality relations for the coefficients
in (\ref{eq:u1},\ref{eq:u2},\ref{eq:u3}), 
\begin{equation}
\sum_{j=1}^{q}c_{\mu,j}c_{\mu',j}^{*}=\delta_{\mu,\mu'},
\end{equation}
thus \eqref{eq:rGmm} becomes 
\begin{multline}
\Gamma_{\mu,\mu'}^{(r)}=c_{\mu,1}^{*}c_{\mu',1}\Bigl\{ c_{\mu,2}c_{\mu',2}^{*}\left(\tfrac{\lambda_{2}}{\lambda_{1}}\right)^{r}+\\
+\left(\delta_{\mu,\mu'}-c_{\mu,1}c_{\mu',1}^{*}-c_{\mu,2}c_{\mu',2}^{*}\right)\left(\tfrac{\lambda_{3}}{\lambda_{1}}\right)^{r}\Bigr\}.\label{eq:Gmm-s}
\end{multline}
From \eqref{eq:Gmm-s}, and after some algebraic
manipulation, we can find the pair correlation functions in terms
of magnetization and transfer matrix eigenvalues, which has the following
form 
\begin{alignat}{1}
\Gamma_{1,1}^{(r)} & =m_{1}(1-m_{1})\left(\tfrac{\lambda_{2}}{\lambda_{1}}\right)^{r},\label{eq:Gm_11}\\
\Gamma_{\mu,\mu}^{(r)} & =(1-m_{1})\left[m_{1}\left(\tfrac{\lambda_{2}}{\lambda_{1}}\right)^{r}+(q-2)\left(\tfrac{\lambda_{3}}{\lambda_{1}}\right)^{r}\right],\label{eq:Gm_mm}\\
\Gamma_{1,\mu}^{(r)} & =-\tfrac{m_{1}(1-m_{1})}{q-1}\left(\tfrac{\lambda_{2}}{\lambda_{1}}\right)^{r},\label{eq:Gm_1mu}\\
\Gamma_{\mu,\mu'}^{(r)} & =(1-m_{1})\left[m_{1}\left(\tfrac{\lambda_{2}}{\lambda_{1}}\right)^{r}-\left(\tfrac{\lambda_{3}}{\lambda_{1}}\right)^{r}\right],\label{eq:Gm_mm'}
\end{alignat}
where $\mu,\mu'=2,\ldots q$ and $\mu\neq\mu'$. An
alternative expression of eqs.(\ref{eq:Gm_11}-\ref{eq:Gm_mm'}) are
given in appendix A.

\begin{figure}[h]
\includegraphics[scale=0.65]{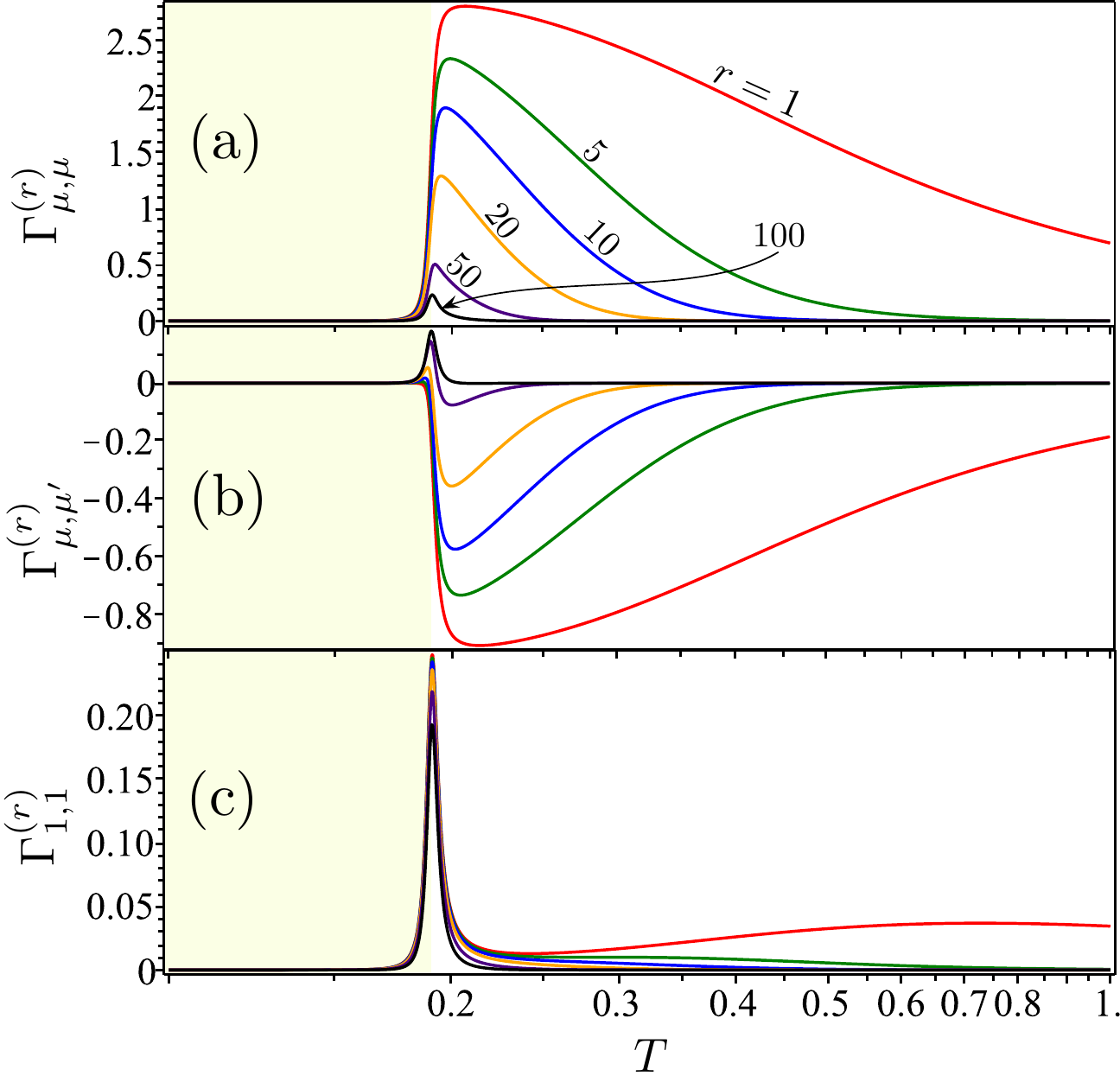}
\caption{\label{fig:The-pair-distribution}The pair distribution correlation
function $\Gamma_{\mu,\mu'}^{(r)}$ as a function of temperature $T$
in logarithmic scale, assuming fixed $q=5$, $J=0.9$, $K=1$, $h=-0.995$
and several values of $r$. (a) For $\mu=\mu'$, with $\mu=\{2,3,\dots\}$.
(b) For $\mu\protect\ne\mu'$ with $\mu,\mu'=\{2,3,\dots\}$. (c)
For $\mu=\mu'=1$.}
\end{figure}

In Fig.~\ref{fig:The-pair-distribution} we illustrate
the pair correlation function $\Gamma_{\mu,\mu'}^{(r)}$ as a function
of temperature in logarithmic scale, here we consider the following
fixed parameters $q=5$, $J=0.9$, $K=1$, $h=-0.995$ and several
distances $r=\{1,5,10,20,50,100\}$. In panel (a) we report for the
case of $\mu=\mu'$ and $\mu=\{2,3,\dots\}$, given by eq.\eqref{eq:Gm_mm}.
It is worth notice that for $T<T_{p}$ (enough below) this amount
is closely null, meaning that the pair distribution spins orientations
have no almost relationship, although for $T>T_{p}$ (enough above)
the system exhibits clearly the relationship between pair spins orientations
and as expected decreases with the distances as well as temperature.
In panel (b) is depicted the correlation function for the case $\mu\ne\mu'$
and $\mu,\mu'=\{2,3,4,\dots\}$, given by eq.\eqref{eq:Gm_mm'}. For
$T<T_{p}$ (enough below) the $\Gamma_{\mu,\mu'}^{(r)}$ become nearly
null correlation function, while for $T>T_{p}$ (enough above) the
systems exhibits a qualitatively different behavior, which in module
decreases with with $r$ and $T$. In panel (c) is displayed for the
case of $\mu=\mu'=1$ (according eq.\eqref{eq:Gm_11}), for this particular
case we observe that the correlation function becomes almost null
far enough from the pseudo-critical temperature $T_{p}$, while at
$T_{p}$ the correlation function illustrate a peak, which decreases
as expected with $r$. The case $\mu=1$ and $\mu'=\{2,3,\dots\}$
given by eq.\eqref{eq:Gm_mm'} is simply the same to the case eq.\eqref{eq:Gm_11}
divided by $(q-1)$.

We can also notice that the four expression given
by (\ref{eq:Gm_11}-\ref{eq:Gm_mm'}) satisfy the following couple
of identities

\begin{alignat}{1}
\Gamma_{1,1}^{(r)}+(q-1)\Gamma_{1,\mu}^{(r)}= & 0,\label{eq:Gr-1st}\\
\Gamma_{\mu,\mu}^{(r)}+\Gamma_{1,\mu}^{(r)}+(q-2)\Gamma_{\mu,\mu'}^{(r)}= & 0.\label{eq:Gr-2nd}
\end{alignat}
An equivalent expressions of the relations (\ref{eq:Gr-1st}-\ref{eq:Gr-2nd})
are given in appendix A.

Similarly, we can obtain a couple of identities for $g$'s, by using
\eqref{eq:Gr-1st} and \eqref{eq:Gr-2nd} which reduce to the following
relations

\begin{alignat}{1}
g_{1,1}^{(r)}+\left(q-1\right)g_{1,\mu}^{(r)}= & m_{1},\label{eq:gr-1st}\\
g_{\mu,\mu}^{(r)}+g_{1,\mu}^{(r)}+(q-2)g_{\mu,\mu'}^{(r)}= & m_{\mu},\label{eq:gr-2nd}
\end{alignat}
which is useful to studying the phase diagram, like
illustrated in Table~\ref{tab:mgs1}.

\begin{table*}[htbp]
\caption{\label{tab:mgs1}The nearest neighbour pair distribution functions 
$g_{\mu,\mu'}^{(1)}$, $\mu,\mu'=2,\ldots q$, magnetization $m_{1}$, and entropy 
$\mathcal{S}$ of the frustrated Potts chain at zero temperature. 
Here $q_{1}=\sqrt{4q-3}$, $q_{2}=\sqrt{(q-1)(q+3)}$, $q_{3}=1+\sqrt{q-1}$,
and $q_{4}=\sqrt{(q-1)^{2}+4}$.}
\begin{ruledtabular}
\begin{tabular}{lcccccc}
GS phase & $g_{1,1}^{(1)}$ & $g_{\mu,\mu}^{(1)}$ & $g_{1,\mu}^{(1)}$ & $g_{\mu,\mu'}^{(1)}$ & $m_{1}$ & $\mathcal{S}$\tabularnewline[0.85em]
\hline 
$FM_{1}$ & $1$ & $0$ & $0$ & $0$ & $1$ & $0$\tabularnewline[0.85em]
$FM_{2}$ & $0$ & $\dfrac{1}{q-1}$ & $0$ & $0$ & $0$ & $0$\tabularnewline[0.85em]
$FR_{1}$ & $0$ & $0$ & $\dfrac{1}{2\left(q-1\right)}$ & $0$ & $\dfrac{1}{2}$ & $\frac{1}{2}\ln\left(q-1\right)$\tabularnewline[0.85em]
$FR_{2}$ & $0$ & $0$ & $0$ & $\dfrac{1}{\left(q-1\right)\left(q-2\right)}$ & 0 & $\ln\left(q-2\right)$\tabularnewline[0.85em]
$FR_{1}$-$FM_{1}$ & $\dfrac{1}{q_{1}}$ & $0$ & $\dfrac{q_{1}-1}{2q_{1}\left(q-1\right)}$ & $0$ & $\dfrac{2q-1+q_{1}}{q_{1}\left(1+q_{1}\right)}$ & $\ln\dfrac{1+q_{1}}{2}$\tabularnewline[0.85em]
$FR_{2}$-$FM_{2}$ & $0$ & $\dfrac{1}{\left(q-1\right)^{2}}$ & $0$ & $\dfrac{1}{\left(q-1\right)^{2}}$ & $0$ & $\ln\left(q-1\right)$\tabularnewline[0.85em]
$FR_{1}$-$FR_{2}$ & $0$ & $0$ & $\dfrac{1}{q\left(q-1\right)}$ & $\dfrac{1}{q\left(q-1\right)}$ & $\dfrac{1}{q}$ & $\ln\left(q-1\right)$\tabularnewline[0.85em]
$FR_{1}$-$FM_{2}$ ($K<0$) & $0$ & $\dfrac{1}{q_{1}\left(q-1\right)}$ & $\dfrac{q_{1}-1}{2q_{1}\left(q-1\right)}$ & $0$ & $\dfrac{2q-2}{q_{1}\left(1+q_{1}\right)}$ & $\ln\dfrac{1+q_{1}}{2}$\tabularnewline[0.85em]
$FR_{2}$-$FM_{1}$ ($K>0$) & $0$ & $0$ & $0$ & $\dfrac{1}{\left(q-1\right)\left(q-2\right)}$ & $0$ & $\ln\left(q-2\right)$\tabularnewline[0.85em]
$FM_{1}$-$FM_{2}$ ($K>0$) & $0$ & $\dfrac{1}{q-1}$ & $0$ & $0$ & $0$ & $0$\tabularnewline[0.85em]
$FM_{1}$-$FM_{2}$ ($K=0$) & $\dfrac{1}{q}$ & $\dfrac{1}{q}$ & $0$ & $0$ & $\dfrac{1}{q}$ & $0$\tabularnewline[0.85em]
$FM_{1}$-$FM_{2}$ ($K<0$) & $\dfrac{1}{2}$ & $\dfrac{1}{2\left(q-1\right)}$ & $0$ & $0$ & $\dfrac{1}{2}$ & $0$\tabularnewline[0.85em]
$Q_{1}$ & $0$ & $\dfrac{1}{q_{2}\left(q-1\right)}$ & $\dfrac{2}{q_{2}\left(q-1+q_{2}\right)}$ & $\dfrac{1}{q_{2}\left(q-1\right)}$ & $\dfrac{2}{q+3+q_{2}}$ & $\ln\dfrac{q-1+q_{2}}{2}$\tabularnewline[0.85em]
$Q_{2}$ & $\dfrac{1}{2q_{3}}$ & $\dfrac{1}{2q_{3}\left(q-1\right)}$ & $\dfrac{1}{2q_{3}\sqrt{q-1}}$ & $0$ & $\dfrac{1}{2}$ & $\ln q_{3}$\tabularnewline[0.85em]
$P_{1}$ & $\dfrac{3-q+q_{4}}{q_{4}\left(q-1+q_{4}\right)}$ & $0$ & $\dfrac{2}{q_{4}\left(q-1+q_{4}\right)}$ & $\dfrac{q-3+q_{4}}{q_{4}\left(q-1\right)\left(q-1+q_{4}\right)}$ & $\dfrac{q+1+q_{4}}{q_{4}\left(q-1+q_{4}\right)}$ & $\ln\dfrac{q-1+q_{4}}{2}$\tabularnewline[0.85em]
$P_{2}$ & $0$ & $\dfrac{1}{\left(q-1\right)^{2}}$ & $0$ & $\dfrac{1}{\left(q-1\right)^{2}}$ & $0$ & $\ln\left(q-1\right)$\tabularnewline[0.85em]
$S$ & $\dfrac{1}{q^{2}}$ & $\dfrac{1}{q^{2}}$ & $\dfrac{1}{q^{2}}$ & $\dfrac{1}{q^{2}}$ & $\dfrac{1}{q}$ & $\ln q$\tabularnewline[0.85em]
\end{tabular}
\end{ruledtabular}
\end{table*}

\subsection{Correlation length}

From transfer matrix eigenvalues, one can observe that $1>\tfrac{\lambda_{2}}{\lambda_{1}}>\tfrac{\lambda_{3}}{\lambda_{1}}$,
thus for $r>1$, we can ignore $\left(\tfrac{\lambda_{3}}{\lambda_{1}}\right)^{r}\rightarrow0$,
since $\left(\tfrac{\lambda_{2}}{\lambda_{1}}\right)^{r}\gg\left(\tfrac{\lambda_{3}}{\lambda_{1}}\right)^{r}$.
Consequently, we can define the correlation length as follows

\begin{equation}
\xi=\left[\ln\left(\frac{\lambda_{1}}{\lambda_{2}}\right)\right]^{-1}.
\end{equation}

An approximation for the correlation length 
in the vicinity of the pseudo-transition point 
immediately follows from equations~\eqref{eq:lam12-appr} 
for the eigenvalues:
\begin{equation}
	\xi \approx \frac{1}{\sqrt{a^2\tau^2+b^2}}.
	\label{eq:ksi-appr}
\end{equation}
At the pseudo-transition point the correlation length reaches 
extremely high values 
\begin{equation}
	\xi_p = b^{-1},
\end{equation}
since the condition $b\ll1$ is necessarily satisfied. 
Indeed, using the expressions of $b$ in the given limit cases, 
we obtain
\begin{equation}
	\xi_p = 
	\left\{
	\begin{array}{ll}
		\tfrac{1}{2}\sqrt{\tfrac{(q-2)x_p k_p}{q-1}} \propto e^{\frac{K-|J|}{2T_p}}, 
		& qFR_2{-}qFM_1;\\[1em]
		\tfrac{x_p}{2}\sqrt{\tfrac{k_p}{q-1}} \propto e^{\frac{K+2J}{2T_p}}, 
		& qFM_2{-}qFM_1.
	\end{array}
	\right.
\end{equation}
The half-width at half-maximum for the peak~\eqref{eq:ksi-appr} 
of the correlation length is $\tilde{\Psi}_{\tau}=\sqrt{3}b/a$ and 
only by numeric factor differs from $\Psi_{\tau}$.

\begin{figure}
 \includegraphics[width=0.23\textwidth]{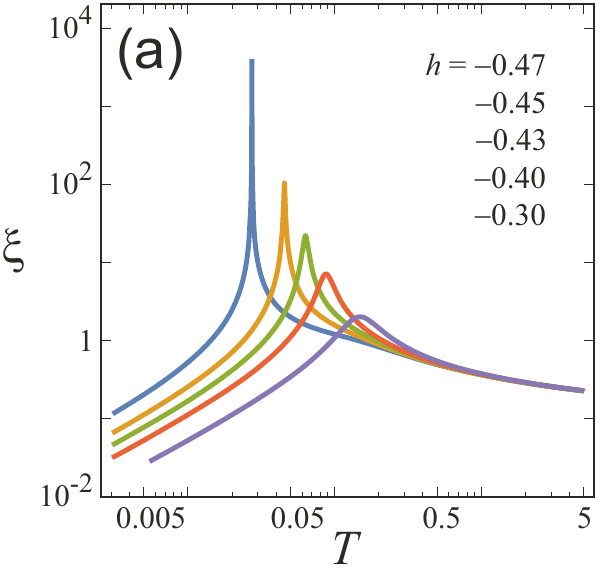} \includegraphics[width=0.23\textwidth]{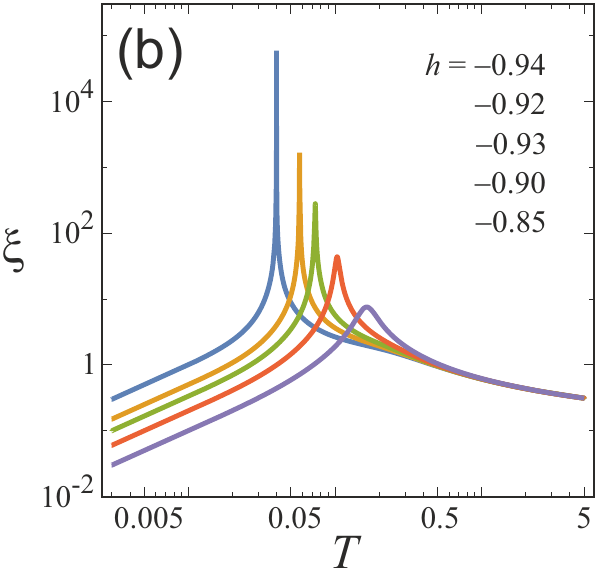} \\
 \includegraphics[width=0.23\textwidth]{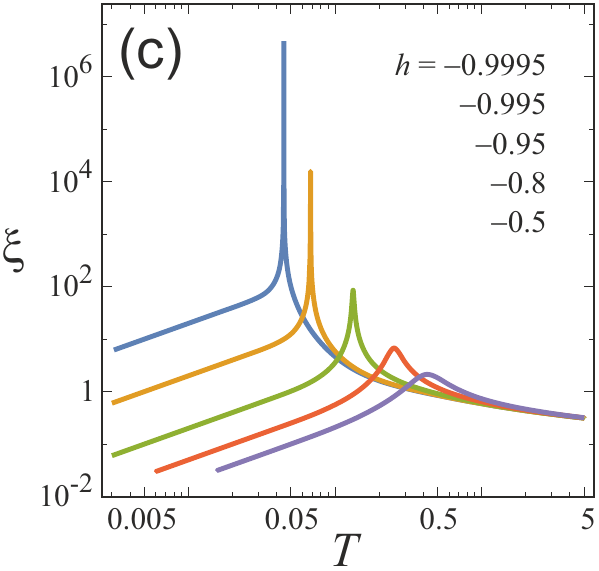} \includegraphics[width=0.23\textwidth]{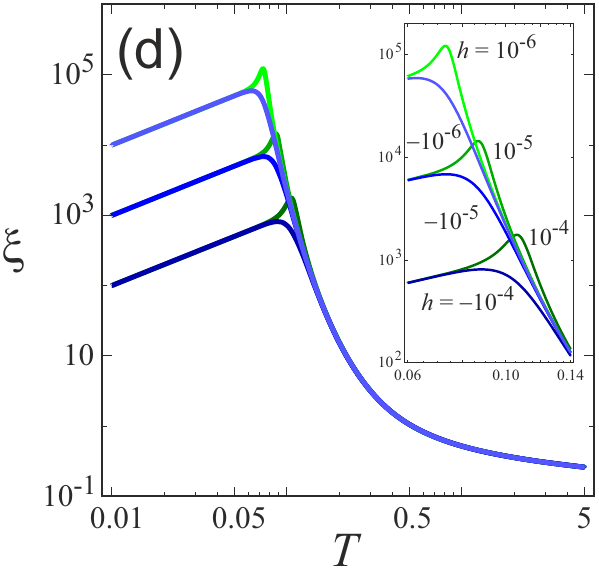}
\caption{The correlation length for  $qFM_{1}$ states with 
(a) $q=5$, $J=-0.5$, $K=1$,
(b) $q=3$, $J=-0.05$, $K=1$,
(c) $q=5$, $J=0.25$, $K=1$,
and for (d) $qFM_{1}$ and $qFM_{2}$ states with $q=10$, $J=1$, $K=0$,
at a given $h$.
}
\label{fig:corrL}
\end{figure}

Figure~\ref{fig:corrL} shows the temperature dependences of the correlation length 
for the same sets of states as in 
Fig.~\ref{fig:mchiFRFM}b, \ref{fig:mchiFRFM}d, \ref{fig:mchiFMM}b, and \ref{fig:mchiFM1FM2}b.
As for the specific heat and susceptibility, the correlation length near the pseudo-transition 
point in Fig.~\ref{fig:corrL}a,b,c shows the giants peaks, which are qualitatively different from 
other cases like shown in Fig.~\ref{fig:corrL}d.

\section{\label{sec:critical_exp}Pseudo-critical exponents}

Now let us analyze the nature of the peaks 
of the specific heat, susceptibility, and correlation length 
around $T_{p}$, whether follow some critical exponent universality. 
A general technique to calculate the critical exponents in the systems 
that exhibit was developed in Ref.~\cite{unv-cr-exp}. 
For the one-dimensional frustrated $q$-state Potts model, we can find 
these quantities directly from approximations~\eqref{eq:C-appr}, 
\eqref{eq:chi1-appr}, and~\eqref{eq:ksi-appr}. 
Considering the region of $\tau$ where the curvature of the peaks becomes positive, 
$b/a<|\tau|\ll T_p/|J|$, we obtain the following asymptotics 
\begin{alignat}{2}
\xi =& \; c_\xi |\tau|^{-1}, &&\qquad c_\xi=a^{-1},\label{eq:xi_asympt}\\
C =& \; c_f |\tau|^{-3}, &&\qquad c_f=\frac{1}{2}a^{-1}b^2,\label{eq:C_asympt}\\
\chi_1 =& \; c_\chi |\tau|^{-3}, &&\qquad c_\chi=\frac{1}{2T_p}a^{-3}b^2.\label{eq:chi_asympt}
\end{alignat}
These found critical exponents are the same as for one-dimensional models 
of the general class with pseudo-transitions~\cite{unv-cr-exp}. 
Combining the critical amplitudes, one can write the following relation
\begin{equation}
	\frac{c_f}{c_\chi}\,c_\xi^2 = T_p,
\end{equation}
which is fulfilled for all pseudo-transitions in the frustrated Potts model.

\begin{figure}[h]
\includegraphics[scale=0.59]{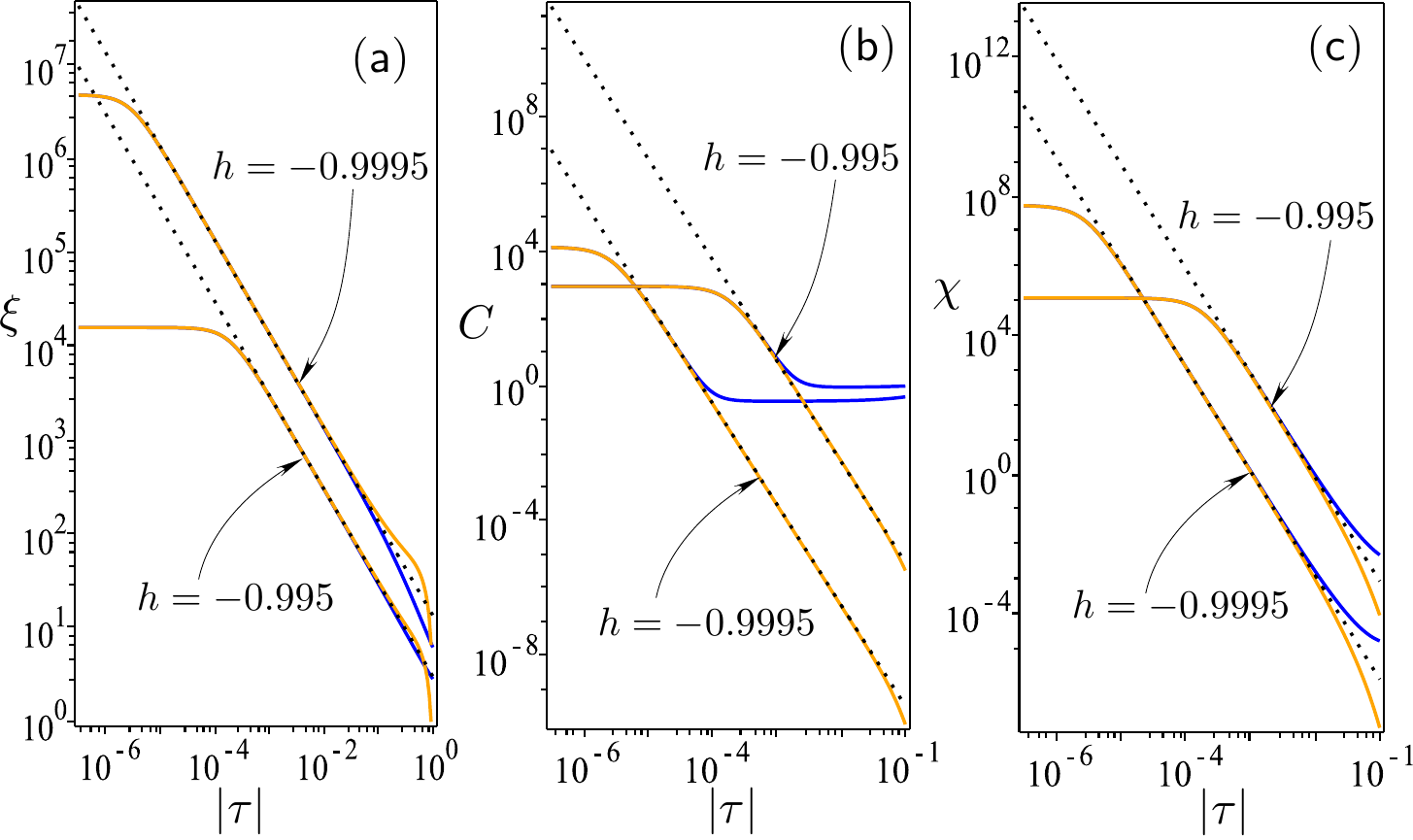}
\caption{\label{fig:crit-exp}(a) The correlation as a function of $|\tau|$ in logarithmic scale, for $q=5$, $J=0.25$, $K=1$, blue curves correspond
to $\tau>0$, orange curves denote the case $\tau<0$, and dotted
lines correspond to $\xi(\tau)$ in asymptotic limit. (b) The specific
heat as a function of $|\tau|$ in logarithmic scale for the same
set of fixed parameters in (a). (c) The magnetic susceptibility as
a function of $|\tau|$ in logarithmic scale, for the parameters assumed
in panel (a).}
\end{figure}

In Fig.~\ref{fig:crit-exp} we verify the power law behavior around
the peak for some physical quantities, these results are only valid
for the ascending and descending part of the peak, while this approach
fails around the top of the peak. In panel (a) we report the correlation
length as a function of $|\tau|$ in logarithmic scale assuming the
parameters $q=5$, $J=0.25$, $K=1$. Blue curves correspond for $\tau>0$,
orange curves denote for $\tau<0$, and dotted lines describe the
asymptotic behavior given by eq.~\eqref{eq:xi_asympt}. We consider
two values of the magnetic field as indicated in Fig.~\ref{fig:crit-exp}a.
For $h=-0.995$, we can observe clearly a straight line with pseudo-critical
exponent $\nu=1$ in a range of $10^{-4}\lesssim\tau\lesssim10^{-1}$
(blue curve). Suchlike behavior we also observe for the case $-10^{-1}\lesssim\tau\lesssim-10^{-4}$
(orange curve). Note that for $|\tau|\lesssim10^{-4}$ the asymptotic
behavior fails because it corresponds to the peak of the curve. Similar
behavior is illustrated for $h=-0.9995$ the pseudo-critical exponent
is accurately described by a straight line with the same exponent
$\nu=1$. Although the asymptotic approach is valid roughly in the
interval $10^{-1}\lesssim|\tau|\lesssim10^{-6}$. In panel (b) is
depicted the specific heat as a function of $|\tau|$ in logarithmic
scale, assuming the same fixed parameters to that considered in panel
(a), the specific heat also exhibits in the ascending a descending
part of the peak as a straight line, which fits precisely to a straight
dotted line with angular coefficient $\alpha=3$ given by eq.~\eqref{eq:C_asympt},
although evidenced in the shorter interval and for smaller $|\tau|$,
the straight lines fail for $|\tau|\lesssim10^{-5}$ ($h=-0.9995$)
and $|\tau|\lesssim5\cdot10^{-3}$ ($h=-0.995$), because we are dealing
with a peak and not a real singularity. Analogous behavior is observed
for the magnetic susceptibility in panel (c), assuming the same set
of fixed parameters given in panel (a). Once again, we observe clearly
a straight line with angular coefficient $\mu=3$ given by eq.~\eqref{eq:chi_asympt},
which is valid for a wider interval compared to that of specific heat. 

In summary, the pseudo-critical exponents are independent of the magnetic
field. We also conclude that these exponents satisfy the same universality
class found previously for other one-dimensional models.

\section{\label{sec:Conclusion}Conclusion}

Although one-dimensional systems could appear thoroughly studied,
they still surprise us by exhibiting unconventional new physics. There
are some peculiar one-dimensional models, that exhibit the presence
of phase transition at finite temperature, under the condition of
nearest neighbor interaction\cite{Galisova,galisova17,*galisova20,strk-cav,on-strk,psd-Ising,pseudo}.
One-dimensional models of statistical physics are attractive from
both theoretical and experimental points of view, often driven to
develop new methods in order to solve more realistic models. 

Here we investigated more carefully the one-dimensional Potts model
with an external magnetic field and anisotropic interaction that selects
 neighboring sites that are in the spin state 1, by using the transfer
matrix method. The largest and the second largest eigenvalues are
almost degenerate for a given temperature, leading to the arising
of pseudo-transition. The rise of anomalous behavior in this model
lies in the peculiar behavior of the transfer matrix elements, all
transfer matrix elements are positive, but some off-diagonal matrix
in the low-temperature region can be extremely small compared to at
least two diagonal elements. We have analyzed, the present model for
several physical quantities assuming $K=1$. The entropy and magnetization
show a steep function around pseudo-critical temperature, rather similar
to a first-order phase transition, while the correlation length, specific
heat, and magnetic susceptibility exhibit sharp peaks, around pseudo-critical
temperature, resembling a second-order phase transition, albeit there
is no true divergence. A further investigation of the pseudo-critical
exponent satisfy the same class of universality previously identified
for other one-dimensional models, these exponents are: for correlation
length $\nu=1$, specific heat $\alpha=3$ and magnetic susceptibility
$\mu=3$. Whereas for $K=0$ (standard Potts chain), we observe a
qualitatively different behavior, such as in entropy there is no step-like
function, there is no sharp peak in specific heat, while broad peak
arises for magnetic susceptibility. 

It is worthy to mention that the pseudo-transition is quite different
from that true phase transition because there is no jump in the first
derivative of free energy, nor divergence in the second derivative
of free energy. In this sense, it would be fairly relevant to observe
this anomalous property experimentally in chemical compounds.

\begin{acknowledgments}
The work was partly supported by the Ministry of Education and Science of RF, 
project No FEUZ-2020-0054, and Brazilian agency CNPq.
\end{acknowledgments}

\appendix

\section{Some addition relations }

Here we give some additional alternative expressions, which would be useful for further analysis,  thus the eqs. (\ref{eq:Gm_11}-\ref{eq:Gm_mm'})
can be reduced to
\begin{alignat}{1}
\Gamma_{1,1}^{(r)} & =\tfrac{\sin^{2}(2\phi)}{4}\left(\tfrac{\lambda_{2}}{\lambda_{1}}\right)^{r},\\
\Gamma_{\mu,\mu}^{(r)} & =\tfrac{\Gamma_{1,1}^{(r)}}{(q-1)^{2}}+\tfrac{(q-2)\sin^{2}(\phi)}{(q-1)^{2}}\left(\tfrac{\lambda_{3}}{\lambda_{1}}\right)^{r},\\
\Gamma_{1,\mu}^{(r)} & =-\tfrac{\Gamma_{1,1}^{(r)}}{q-1},\\
\Gamma_{\mu,\mu'}^{(r)} & =\tfrac{\Gamma_{1,1}^{(r)}}{(q-1)^{2}}-\tfrac{\sin^{2}(\phi)}{(q-1)^{2}}\left(\tfrac{\lambda_{3}}{\lambda_{1}}\right)^{r}.
\end{alignat}

Considering the identities given by \eqref{eq:Gr-1st} and \eqref{eq:Gr-2nd},
we obtain the following identity which must satisfy: 
\begin{multline}
\Gamma_{1,1}^{(r)}+\left(q-1\right)\Gamma_{\mu,\mu}^{(r)}+2\left(q-1\right)\Gamma_{1,\mu}^{(r)}+{}\\
+\left(q-1\right)\left(q-2\right)\Gamma_{\mu,\mu'}^{(r)}=0.
\end{multline}
An equivalent relation we can verify, so the pair average distribution
functions obey the identity 
\begin{multline}
g_{1,1}^{(r)}+\left(q-1\right)g_{\mu,\mu}^{(r)}+2\left(q-1\right)g_{1,\mu}^{(r)}+{}\\
+\left(q-1\right)\left(q-2\right)g_{\mu,\mu'}^{(r)}=1.
\end{multline}

Below we simplify for the nearest neighbors PDF $g_{\mu,\mu'}^{(1)}$, which 
can be written as 
\begin{alignat}{1}
g_{1,1}^{(1)} & =\tfrac{\left(\lambda_{1}-\lambda_{3}-(q-1)t_{2}\right)\left(\lambda_{1}+\lambda_{2}-\lambda_{3}-(q-1)t_{2}\right)}{\lambda_{1}\left(\lambda_{1}-\lambda_{2}\right)},\label{eq:g11}\\
g_{\mu,\mu}^{(1)} & =\tfrac{\left(\lambda_{3}+t_{2}\right)\left(\lambda_{3}-\lambda_{2}+(q-1)t_{2}\right)}{\left(q-1\right)\lambda_{1}\left(\lambda_{1}-\lambda_{2}\right)},\label{eq:gpp}\\
g_{1,\mu}^{(1)} & =\tfrac{\left(\lambda_{1}-\lambda_{3}-(q-1)t_{2}\right)\left(\lambda_{3}-\lambda_{2}+(q-1)t_{2}\right)}{\left(q-1\right)\lambda_{1}\left(\lambda_{1}-\lambda_{2}\right)},\label{eq:g1p}\\
g_{\mu,\mu'}^{(1)} & =\tfrac{\left(\lambda_{3}-\lambda_{2}+(q-1)t_{2}\right)t_{2}}{\left(q-1\right)\lambda_{1}\left(\lambda_{1}-\lambda_{2}\right)},\label{eq:gppp}
\end{alignat}
where $\mu,\mu'=2,\ldots q$, and $\mu\neq\mu'$. This amounts are useful to analysis the phase boundary properties illustrated in Table~\ref{tab:mgs1}.

\bibliography{apsPotts}

\end{document}